\documentclass[aps,prd,nofootinbib,preprint]{revtex4-1}

\usepackage{graphicx}
\usepackage{hyperref}
\usepackage{amsmath}
\usepackage{breakurl}
\usepackage[caption=false]{subfig}
\usepackage{multirow}

\usepackage{color}

\usepackage[normalem]{ulem} 
\renewcommand\sout{\bgroup \color[rgb]{0.55,0.00,0.99} \ULdepth=-.5ex \ULset}

\newcommand{\Jpsi}{$J/\psi$}
\newcommand{\Ups}{$\Upsilon$(1S)}

\def\figext{eps}
\def\figext{pdf}

\usepackage{times}
\begin{document}


\title{Constraining gluon density of pions at large $x$ by
  pion-induced $J/\psi$ production}

\author{Wen-Chen Chang} 
\affiliation{Institute of Physics, Academia Sinica, Taipei 11529,
  Taiwan}

\author{Jen-Chieh Peng}
\affiliation{Department of Physics, University of Illinois at
  Urbana-Champaign, Urbana, Illinois 61801, USA}

\author{Stephane Platchkov}
\affiliation{IRFU, CEA, Universit\'{e} Paris-Saclay, 91191
  Gif-sur-Yvette, France}

\author{Takahiro Sawada}
\affiliation{Department of Physics, Osaka City University, Osaka 558-8585, Japan}


\begin{abstract}

The gluon distributions of the pion obtained from various global fits
exhibit large variations among them. Within the framework of the color
evaporation model, we show that the existing pion-induced \Jpsi~
production data, usually not included in the global fits, can impose
useful additional constraints on the pion parton distribution
functions (PDFs). In particular, these data can probe the pion's gluon
densities at large $x$. Existing pion-induced \Jpsi~ data covering a
broad range of beam momenta are compared with next-to-leading-order
QCD calculations using various sets of pion PDFs. It is found that
\Jpsi~ data measured at forward rapidity and at sufficiently high beam
momentum are sensitive to the large-$x$ gluon distribution of
pions. The current \Jpsi~ data favor the
Sutton-Martin-Roberts-Stirling and Gluck-Reya-Vogt pion PDFs,
containing significant gluon content at large $x$.

\end{abstract}

\maketitle

\section{Introduction}
\label{sec:introduction}

The pion, as the Goldstone boson of dynamical chiral symmetry breaking
of the strong interaction, is the lightest QCD bound state. Because of
its light mass, the pion plays a dominant role in the long-range
nucleon-nucleon interaction~\cite{Yukawa}. Understanding the pion's
internal structure is important to investigate the low-energy,
nonperturbative aspects of QCD~\cite{Horn:2016rip}. Even though the
pion is theoretically simpler than the proton, its partonic structure
is much less explored. As scattering off a pion target is not
feasible, current knowledge on pion parton distribution functions
(PDFs) mostly relies on the pion-induced Drell-Yan data. Since these
fixed-target data are mostly sensitive to the valence-quark
distributions at $x >0.2$, the sea and gluon densities are essentially
unconstrained.

In principle, the prompt-photon production process $\pi N \rightarrow
\gamma X$ can constrain the gluon content of pions through the $Gq
\rightarrow \gamma q$ subprocess, but the experimental uncertainties
are large. Production of heavy quarkonia, like \Jpsi~ and \Ups, with
the pion beam has distinctive advantages: The cross sections are large
and they can be readily detected via the dimuon decay channel. These
datasets have been shown to be sensitive to both the quark and gluon
distributions of the incident pion with model-dependent assumptions of
quarkonia fragmentation~\cite{Gluck:1977zm,Barger:1980mg}. The
interesting possibility of accessing the pion PDFs from leading
neutron deep inelastic scattering (DIS) data has been considered with
promising results~\cite{Khoze:2006hw, McKenney:2015xis}. However, this
method is subject to large systematic uncertainties, and further
studies on the uncertainties of the pion splitting function and the
off-shellness of virtual pion are
required~\cite{Qin:2017lcd,Perry:2018kok}. To precisely determine the
sea quark content of pions, there was a suggestion of performing the
Drell-Yan measurement with $\pi^+$ and $\pi^-$ beams on the isoscalar
deuterium target~\cite{Londergan:1995wp}, and such a measurement is
planned in a future experiment~\cite{AMBER}.

Until a couple of years ago, knowledge of the pion PDFs was limited to
global analyses carried out more than two decades ago: Owens
(OW)~\cite{Owens:1984zj},
Aurenche-Baier-Fontannaz-Kienzle-Focacci-Werlen
(ABFKW)~\cite{Aurenche:1989sx}, Sutton-Martin-Roberts-Stirling
(SMRS)~\cite{Sutton:1991ay}, and Gluck-Reya-Vogt
(GRV)~\cite{Gluck:1991ey}, and Gluck-Reya-Schienbein
(GRS)~\cite{Gluck:1999xe}. These analyses were based mostly on
pion-induced Drell-Yan, often on prompt-photon and in some cases on
\Jpsi~ production data. New analyses were performed only recently,
using the same Drell-Yan data in Bourrely-Soffer
(BS)~\cite{Bourrely:2018yck} as well as both the Drell-Yan and
direct-photon data in xFitter~\cite{Novikov:2020snp}. The analysis of
JAM~\cite{Barry:2018ort} included both the Drell-Yan data and, for the
first time, the leading neutron tagged electroproduction data. The
experimental situation is also evolving. After more than two decades,
a new measurement of pion-induced Drell-Yan production cross sections
was performed by the CERN COMPASS Collaboration~\cite{COMPASS}. The
data are expected to be available in the near future. A proposal
dedicated to investigating the pion and kaon structure at the future
electron-ion collider in the U.S.~\cite{EIC} was recently described in
Ref.~\cite{Aguilar:2019teb}.


On the theoretical side, the interest in the meson structure has
considerably increased in recent years. Numerous new calculations,
based on the chiral-quark model~\cite{Nam:2012vm,Watanabe:2016lto,
  Watanabe:2017pvl}, Nambu-Jona-Lasinio model~\cite{Hutauruk:2016sug},
light-front Hamiltonian~\cite{Lan:2019vui, Lan:2019rba}, holographic
QCD~\cite{deTeramond:2018ecg, Watanabe:2019zny}, maximum entropy
method~\cite{Han:2018wsw}, and continuum functional approach using
Dyson-Schwinger equations (DSE)~\cite{Chang:2014lva, Chang:2014gga,
  Chen:2016sno, Shi:2018mcb, Bednar:2018mtf, Ding:2019lwe}, became
available. A major breakthrough in lattice QCD~\cite{Ji:2013dva} led
several groups to perform a direct calculation of the pion valence $x$
distribution~\cite{Chen:2018fwa, Sufian:2019bol, Izubuchi:2019lyk,
  Joo:2019bzr, Sufian:2020vzb}. Further improvement in the accuracy of
the lattice calculations is anticipated. As of today, most of the
theoretical predictions deal with the pion valence-quark distribution
only. The gluon and sea PDFs are predicted solely within the DSE
continuum approach~\cite{Ding:2019lwe}.

In this work we investigate the sensitivity of the \Jpsi~ production
data to the pion PDFs. The theoretical challenge of this reaction
comes from the treatment of the hadronization of $c \bar{c}$ pairs
into a charmonium bound state. This nonperturbative process has been
modeled in several theoretical approaches including the color
evaporation model (CEM)~\cite{CEM}, the color-singlet model
(CSM)~\cite{CSM}, and nonrelativistic QCD (NRQCD)~\cite{NRQCD}. The
CEM assumes a constant probability for $c \bar{c}$ pairs to hadronize
into a given charmonium. In the CSM, the production of \Jpsi~ is
assumed to be through the color-singlet $c \bar{c}$ channel of the
same quantum numbers as \Jpsi. The NRQCD expands the calculations by
the powers of the average velocity of $c \bar{c}$ pairs in the rest
frame of \Jpsi. The hadronization probability of each $c \bar{c}$ pair
depends on its color and spin state. More details about these
theoretical frameworks can be found in Ref.~\cite{Schuler}. In
general, the CSM and NRQCD provide a good description of data taken at
collider energies but fail to explain measurements at fixed-target
energies~\cite{Maltoni:2006yp}.

To explore the constraints on the pion PDFs by the \Jpsi~ production
process, a theoretical model with a minimal number of parameters is
preferred. A great feature of the CEM is that it is essentially
parameter-free, except for a single effective parameter that accounts
for the probability of $c \bar c$ pairs to hadronize into a particular
quarkonium bound state. In spite of its well-known
limitations~\cite{Bodwin:2005hm}, the CEM gives a good account of many
features of fixed-target \Jpsi~ cross section data with proton beams,
including their longitudinal momentum ($x_F$)
distributions~\cite{Gavai:1994in, Schuler:1996ku} and the collider
data at RHIC, Tevatron, and LHC~\cite{Nelson:2012bc,
  Lansberg:2020rft}. Since the proton PDFs are well known from other
processes (DIS, Drell-Yan, etc.), the proton-induced \Jpsi~ data are
useful for validating the CEM as a suitable model. In contrast, the
pion-induced \Jpsi~ data involve the poorly known pion's PDFs.
Therefore, we use the CEM to study the sensitivity of available \Jpsi~
production data to the pion PDFs, especially the gluon distributions.

In the fixed-target energy domain, where the transverse momentum of
the charmonium is less than its mass, the charmonium production is
dominated by the quark-antiquark ($q \bar{q}$) and gluon-gluon fusion
($GG$) partonic processes. The shape of the longitudinal momentum
$x_F$ cross section is, therefore, sensitive to the quark and gluon
parton distributions of colliding hadrons. Since the nucleon PDFs are
known with good accuracy, the measurement of the $x_F$ distribution of
\Jpsi~ production with the pion beam provides, within the theoretical
model uncertainties, valuable information about the pion quark and
gluon partonic distributions. Our study is performed using
next-to-leading-order (NLO) CEM calculation, including the recent
nucleon PDFs. The available pion-induced \Jpsi~ data on hydrogen and
several light-mass nuclear targets are compared to calculations using
the available pion PDFs. Over the broad energy range considered, all
pion PDF sets provide reasonable agreement with the $x_F$-integrated
cross sections. In contrast, for the $x_F$ distributions, we find that
the agreement between data and calculations strongly depends on the
magnitude and shape of the pion gluon distribution.

This paper is organized as follows. In Sec.~\ref{sec:CEM}, we briefly
describe the CEM framework for the calculations of \Jpsi~ production
cross sections in the collisions of pions and nucleons. Some
distinctive features of parton densities in various pion PDFs used for
the calculations are presented in Sec.~\ref{sec:PDFs}. The NLO CEM
calculations using various pion PDFs are compared with the existing
\Jpsi~ production data in
Sec.~\ref{sec:results}. Section~\ref{sec:syst} shows the results of
systematic study for the CEM calculation. We discuss our findings from
the comparison of CEM calculations with data in
Sec.~\ref{sec:discussion}, followed by a summary in
Sec.~\ref{sec:summary}.

\section{Color Evaporation Model and Heavy-Quark Pair Production}
\label{sec:CEM}

The theoretical treatment of heavy quarkonium production consists of
the QCD description of the production of heavy-quark pairs ($Q
\bar{Q}$) at the parton level, and their subsequent hadronization into
the quarkonium states. One of the theoretical approaches is
NRQCD~\cite{NRQCD}, where the cross section of quarkonium production
is expanded in terms of the strong coupling constant $\alpha_S$ and
the $Q \bar{Q}$ velocity. The cross section is factorized into the
hard and soft parts for each color and spin state of the $Q \bar{Q}$
pairs. The short-distance hard part is calculated perturbatively as a
series of $\alpha_S$ in perturbative QCD (pQCD). The soft part
consists of long-distance matrix elements (LDMEs) characterizing the
probability of hadronization process for each color and spin
state. The LDMEs are determined by a fit to the experimental data. In
the color-singlet model~\cite{CSM}, the production channel is assumed
to be the color-singlet $Q \bar{Q}$ state with quantum numbers exactly
matching those of the heavy quarkonium.

Based on quark-hadron duality, the CEM assumes a constant probability
for $Q \bar{Q}$ pairs to hadronize into a quarkonium state. Taking
\Jpsi~ as an example, one first produces a $c \bar{c}$ pair via
various QCD hard processes. For $c \bar{c}$ with an invariant mass
$M_{c\bar{c}}$ less than the $D \bar{D}$ threshold, a constant
probability $F$, specific for each quarkonium, accounts for the
hadronization of $c \bar{c}$ pairs into the colorless \Jpsi~ state.

In the CEM, the differential cross section $d\sigma/dx_F$ for \Jpsi~
from the $\pi N$ collision is expressed as
\begin{align}
\frac{d\sigma}{dx_F}|_{J/\psi}=& F \sum\limits_{i,j=q, \bar{q},
  G} \int_{2 m_c} ^{2 m_{D}} dM_{c \bar{c}} \frac{2M_{c \bar{c}}}{s\sqrt{x_F^2+4{M_{c \bar{c}}}^2/s}} \nonumber \\
 \times f^{\pi}_{i}(x_1, \mu_{F}) & f^{N}_{j}(x_2, \mu_{F}) \hat{\sigma}[ij \rightarrow c \bar{c} X](x_1 p_{\pi} , x_2 p_{N} , \mu_{F},
\mu_{R}),
\label{eq:eq1}
\end{align}
\begin{align}
  x_F = 2 p_L/\sqrt{s} \mbox{,   } x_{1,2} =  \frac{\sqrt{x_F^2+4{M_{c \bar{c}}}^2/s} \pm x_F}{2} 
\end{align}
where $i$ and $j$ denote the interacting partons (gluons, quarks and
antiquarks) and $m_c$, $m_D$, and $M_{c \bar{c}}$ are the masses of
the charm quark, $D$ meson, and $c \bar{c}$ pair, respectively. The
$f^{\pi}$ and $f^{N}$ are the corresponding pion and nucleon parton
distribution functions, respectively, evaluated at the corresponding
Bjorken-$x$, $x_1$ and $x_2$, at the factorization scale $\mu_F$.

The short-distance differential cross section of heavy-quark pair
production $\hat{\sigma}[ij \rightarrow c \bar{c} X]$ is calculable as
a perturbation series in the strong coupling $\alpha_s(\mu_R)$
evaluated at the renormalization scale $\mu_R$. The variable $s$ is
the square of the center-of-mass energy of the colliding $\pi$-$N$
system, and $p_L$ is the longitudinal momentum of detected dimuon pair
in the center-of-mass frame of $\pi$-$N$. It is assumed that the
momenta of \Jpsi~ and $c \bar{c}$~ are approximately the same.

As mentioned above, the hadronization factor $F$ is assumed to be
universal, independent of the kinematics and the spin state of $c
\bar{c}$ and the production subprocess. Therefore a unique feature of
the CEM calculation is that the relative weight of each subprocess in
$d\sigma/dx_F$, is fixed solely by the convolution of partonic-level
cross sections $\hat{\sigma}$ and associated parton density
distributions $f^{\pi}$ and $f^{N}$ and, in particular, is independent
of the $F$ factor. The $F$ factor is to be determined as the
normalization parameter in the fit to the experimental
measurements. The assumption of a common $F$ factor for different
subprocesses greatly reduces the number of free parameters in the CEM.

The leading-order $[\mathcal{O}(\alpha_S^2)]$ calculations of hard QCD
kernel $\hat{\sigma}[ij \rightarrow c \bar{c} X]$ include the
quark-antiquark ($q \bar{q}$) and gluon-gluon fusion ($GG$)
diagrams. Additional quark-gluon Compton scattering ($Gq$, $G\bar{q}$)
and virtual gluon corrections enter the NLO
$[\mathcal{O}(\alpha_S^3)]$ calculations. The contributing partonic
subprocesses in the fixed-order LO and NLO calculations are listed
explicitly below~\cite{Nason:1987xz}:
\begin{eqnarray}
q + \bar{q} \rightarrow Q + \bar{Q}, &  \alpha_S^2,\alpha_S^3 \nonumber \\
G + G \rightarrow Q + \bar{Q}, &  \alpha_S^2,\alpha_S^3 \nonumber \\
q + \bar{q} \rightarrow Q + \bar{Q} + g, &  \alpha_S^3 \nonumber \\
G + G \rightarrow Q + \bar{Q} + g, &  \alpha_S^3 \nonumber \\
G + q \rightarrow Q + \bar{Q} + q, &  \alpha_S^3 \nonumber \\
G + \bar{q} \rightarrow Q + \bar{Q} + \bar{q}, &  \alpha_S^3 .
\label{eq:eq2}
\end{eqnarray}
Inclusion of both real and virtual gluon emission diagrams is
necessary for calculating the full $\mathcal{O}(\alpha_S^3)$ cross
sections.

In this work, we utilize the theoretical framework of NLO calculation
of the total cross sections for production of the heavy-quark pair,
developed by Nason {\em et al.}~\cite{Nason:1987xz, Nason:1989zy,
  Mangano:1992kq}. This framework has been widely used in the
calculation of heavy-quark production. For example, it has been
adopted in the NLO calculation of the CEM for \Jpsi~ production in
hadronic collisions\cite{Gavai:1994in, Schuler:1996ku,
  Nelson:2012bc}. With a few parameters including the heavy-quark mass
$m_c$ and hadronization factor $F$, the CEM calculations adequately
reproduced the fixed-target data with proton, antiproton and pion
beams~\cite{Gavai:1994in, Schuler:1996ku}, as well as the collider
data~\cite{Nelson:2012bc, Lansberg:2020rft}.


\section{Pion PDFs}
\label{sec:PDFs}

Pion-induced Drell-Yan data are included in all global analyses for
the determination of the pion PDFs. However, Drell-Yan
data~\cite{Badier:1983mj, Anassontzis:1987hk, falciano86, conway}
constrain mainly the valence-quark distribution. Without additional
observables, the sea and gluon distributions remain practically
unknown. Their magnitude can be only inferred through the momentum sum
rule and valence-quark sum rule. Different approaches have been taken
to access the gluon and sea quark distributions: (i) utilizing \Jpsi~
production data in OW~\cite{Owens:1984zj}; (ii) utilizing the
direct-photon production data~\cite{Bonesini:1987mq} in
ABFKW~\cite{Aurenche:1989sx}, SMRS~\cite{Sutton:1991ay},
GRV~\cite{Gluck:1991ey}, and xFitter~\cite{Novikov:2020snp}; (iii)
utilizing leading neutron DIS~\cite{Chekanov:2002pf, Aaron:2010ab} in
JAM~\cite{Barry:2018ort}. In addition, some pion PDFs are based on
theoretical modeling. For example, GRS~\cite{Gluck:1999xe} utilized a
constituent quark model to relate the gluon and antiquark density, and
BS~\cite{Bourrely:2018yck} assumed quantum statistical distributions
for all parton species with an universal temperature. We note that the
OW analysis was performed at LO, whereas a NLO fit was carried out for
all other analyses. Uncertainty bands for the resulting parton density
distributions are available for two most recent global fits, JAM and
xFitter. It was recently shown that the soft-gluon threshold
resummation correction modifies the extraction of valence-quark
distribution and, particularly, its falloff toward
$x=1$~\cite{Aicher:2010cb}. This correction has not been implemented
in any of the pion global analyses yet and it should affect only the
calculated shape at the highest $x_F$ region.

\begin{figure}[htbp]
\centering
\subfloat[]
{\includegraphics[width=0.8\columnwidth]{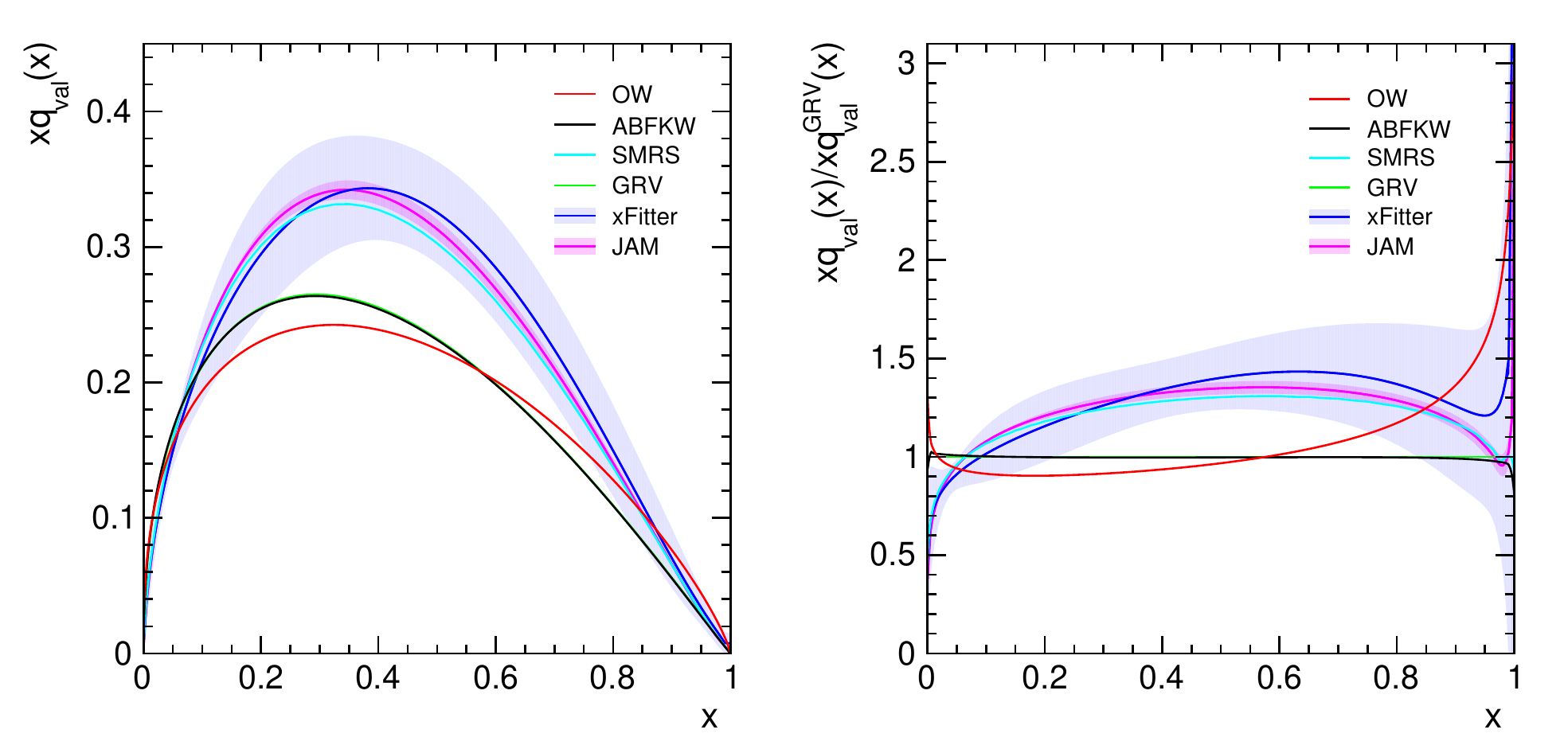}}
\qquad
\subfloat[]
{\includegraphics[width=0.8\columnwidth]{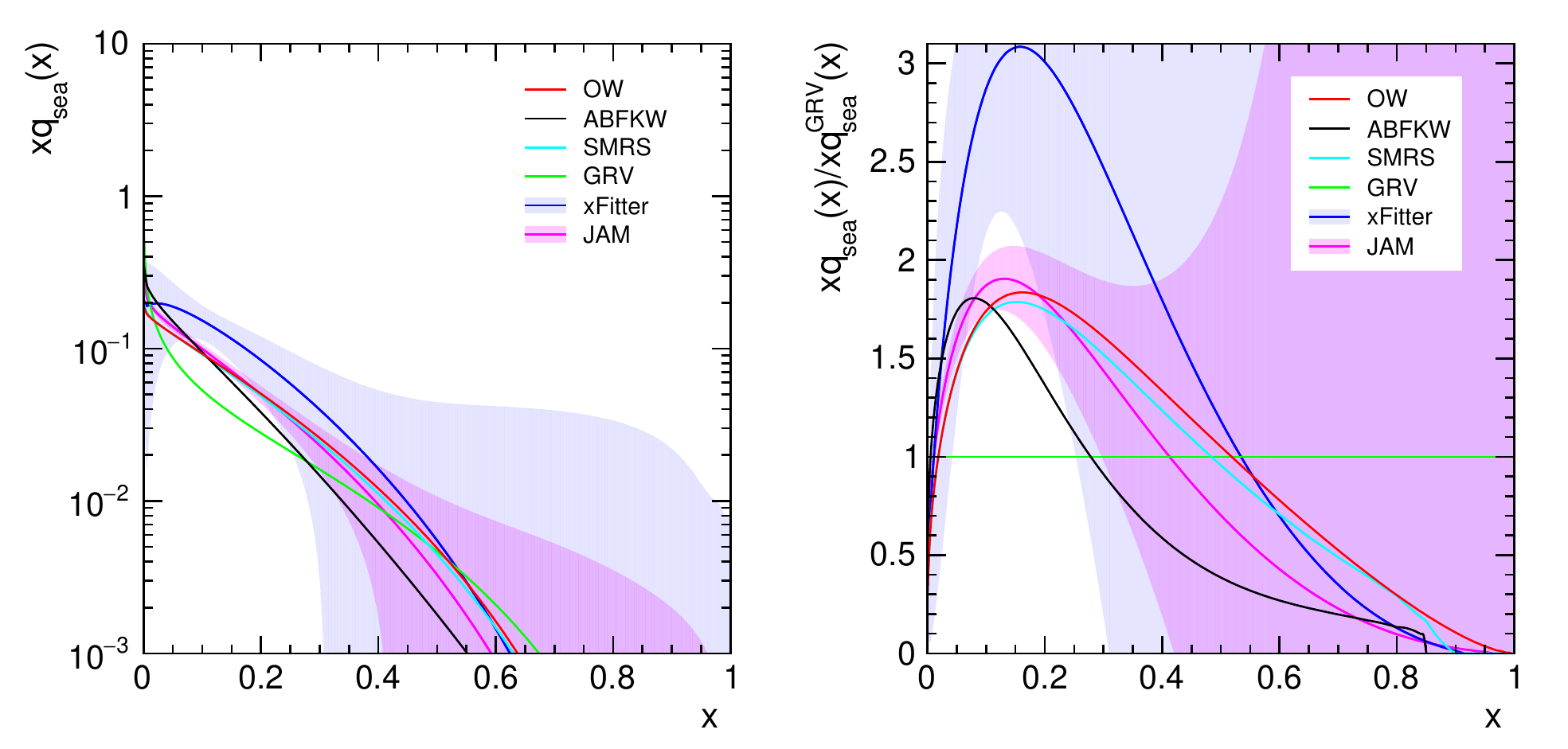}}
\qquad
\subfloat[]
{\includegraphics[width=0.8\columnwidth]{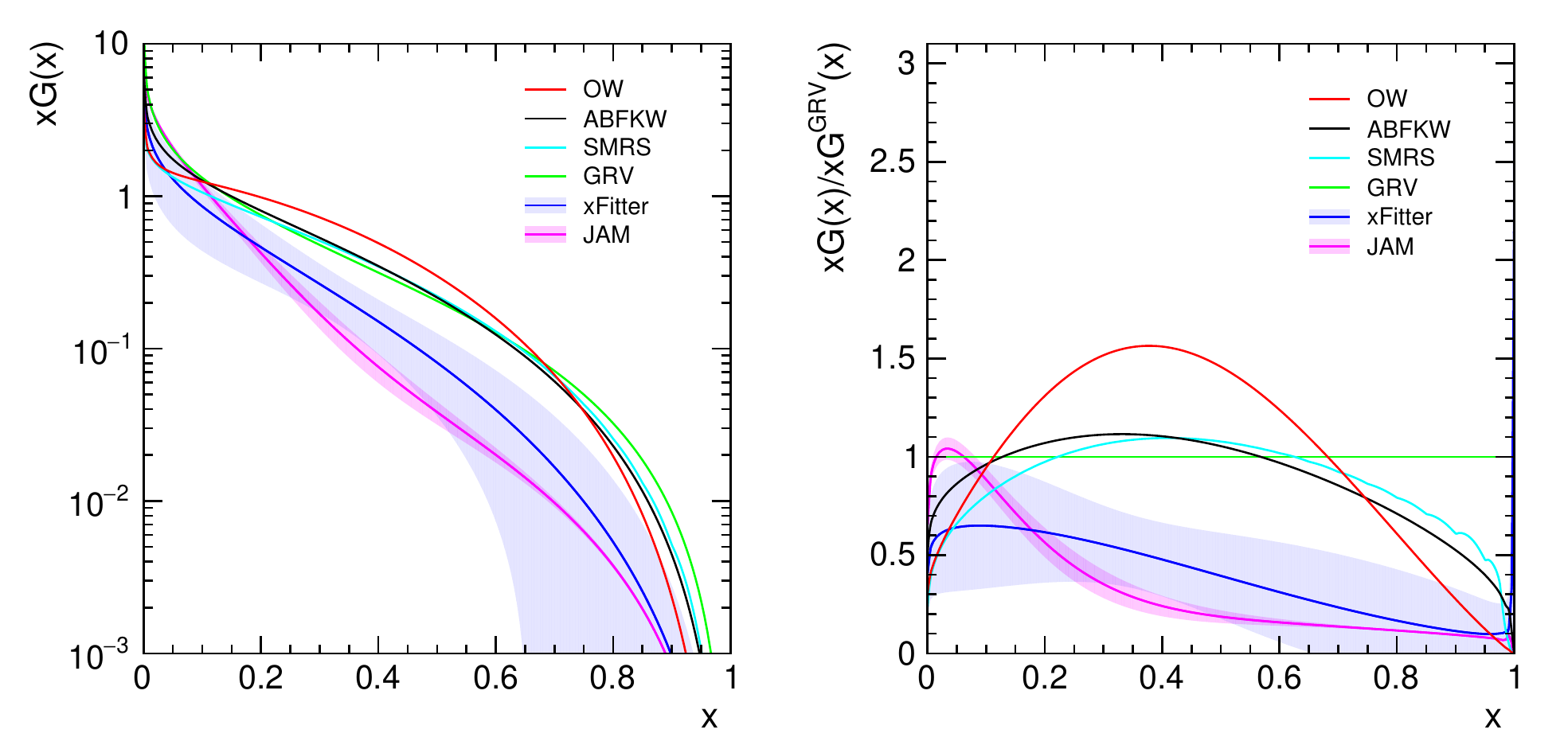}}
\caption
[\protect{}] {Momentum density distributions $[xf(x)]$ of (a) valence
  quarks, (b) sea quarks, and (c) gluons of various pion PDFs and
  their ratios to GRV, at the scale of \Jpsi~ mass ($Q^2$= 9.6
  GeV$^2$).}
\label{fig_pdf}
\end{figure}

Figure~\ref{fig_pdf} compares the valence, sea, and gluon momentum
distributions of the OW, ABFKW, SMRS, GRV, JAM and xFitter pion PDFs
at the scale of \Jpsi~ mass. For clarity, we also show their ratios to
GRV. Within $x \sim$0.1--0.8, the valence-quark distributions of SMRS,
JAM and xFitter are close to each other, whereas those of OW, ABFKW,
and GRV are lower by 20\%--30\%. The sizable error bands of the sea
distributions provided by JAM and xFitter clearly indicate that the
pion sea remains poorly known. As for the gluon distributions, the
early PDF sets of OW, ABFKW, SMRS, and GRV have relatively large
densities for $x>0.1$, at variance with the recent xFitter and JAM
PDFs that lie significantly lower. The spread of the gluon
distributions around $x=0.5$ among these six PDFs is even larger than
the uncertainties of xFitter and JAM PDFs.

Table~\ref{tab:xmoment} lists the momentum fractions of valence quarks
($\bar{u}_{val}$), sea quarks ($\bar{u}_{sea}$), and gluons ($G$) of
negative pions estimated by various pion PDFs at $Q^2$= 9.6 GeV$^2$,
following the definitions of $\bar{u}_{val}(x) =
\bar{u}(x)-\bar{u}_{sea}(x)$, $d_{val}(x)=d(x)-d_{sea}(x)$,
$\bar{u}_{val}(x) = d_{val}(x)$, and
$\bar{u}_{sea}(x)=d_{sea}(x)$. The values for the valence quarks show
differences of up to 15\%--20\% but are nearly equal for the two more
recent PDFs, JAM and xFitter. The gluon first moments vary from 0.29
for xFitter to 0.51 for GRV. The low gluon value in xFitter is
compensated by a much larger sea contribution.

\begin{table}[htbp]   
\centering
\begin{tabular}{|c|c|c|c|c|c|c|c|c|c|}
\hline
PDF
& $\int_0^1 x\bar{u}_{val}(x) dx$ & $\int_0^1 x\bar{u}_{sea}(x) dx$ & $\int_0^1 xG(x) dx$ \\
\hline
OW & 0.203 & 0.026 & 0.487 \\
ABFKW & 0.205 & 0.026 & 0.468 \\
SMRS & 0.245 & 0.026 & 0.394 \\
GRV & 0.199 & 0.020 & 0.513 \\
JAM$^{a}$ & $0.225 \pm 0.003$ & $0.028 \pm 0.002$ & $0.365 \pm 0.016$ \\
xFitter$^{a}$ & $0.228 \pm 0.009$ & $0.040 \pm 0.020$ & $0.291 \pm 0.119$ \\
\hline
\end{tabular}
\caption {Momentum fractions of valence quarks, sea quarks and gluons
  of various pion PDFs for $\pi^-$ at the scale $Q^2$= 9.6
  GeV$^2$.\\ $^{a}$Uncertainties estimated from the member PDF sets.}
\label{tab:xmoment}
\end{table}

\section{Results of NLO CEM Calculations}
\label{sec:results}

In this section, we explore the sensitivity of the NLO CEM
calculations to the various global fit parametrizations of the pion
PDFs. We select four of them, namely, SMRS and GRV, as the most widely
used for a long time, and the two most recent fits, xFitter and
JAM. Out of the three possible parametrizations for SMRS, we choose
the one in which the sea quarks carry 15\% of the pion momentum at
$Q^2$= 4 GeV$^2$. As illustrated in Fig.~\ref{fig_pdf}, SMRS, JAM, and
xFitter have quite similar valence-quark distributions while the
magnitude of the GRV distribution is lower, by up to 20\%--30\%. As for
the gluon distributions, SMRS and GRV have similar shapes and
magnitudes, while the magnitudes of xFitter and JAM are significantly
smaller, by a factor of 2--4.

As a first step, we compare the NLO CEM cross sections integrated over
$x_F > 0$ for the process $\pi^- N \rightarrow J/\psi X$ for each of
the four pion PDFs with the available measurements as a function of
the center-of-mass energy $\sqrt s$ of the reaction. The calculations
are performed using the nucleon CT14nlo PDFs~\cite{CT14nlo} under the
LHAPDF framework~\cite{LHAPDF5, LHAPDF6}. The cross sections are
evaluated with a charm quark mass $m_c= 1.5$ GeV/$c^2$ and
renormalization and factorization scales of $\mu_R= m_c$ and $\mu_F =
2 m_c$, respectively~\cite{Mangano:1992kq}. The experimental cross
sections are taken from the compilation of Ref.~\cite{Schuler}. For
the sake of completeness, the subsequent measurement from the WA92
experiment~\cite{Alexandrov:1999ch} is also included, after correcting
it for the nuclear dependence. The hadronization factors $F$ are
assumed to be energy independent and are determined by the best fit to
the data for the central values of each pion PDF. The uncertainties of
xFitter and JAM PDFs are not taken into account here.

\begin{figure}[htbp]
\centering
\includegraphics[width=0.9\columnwidth]{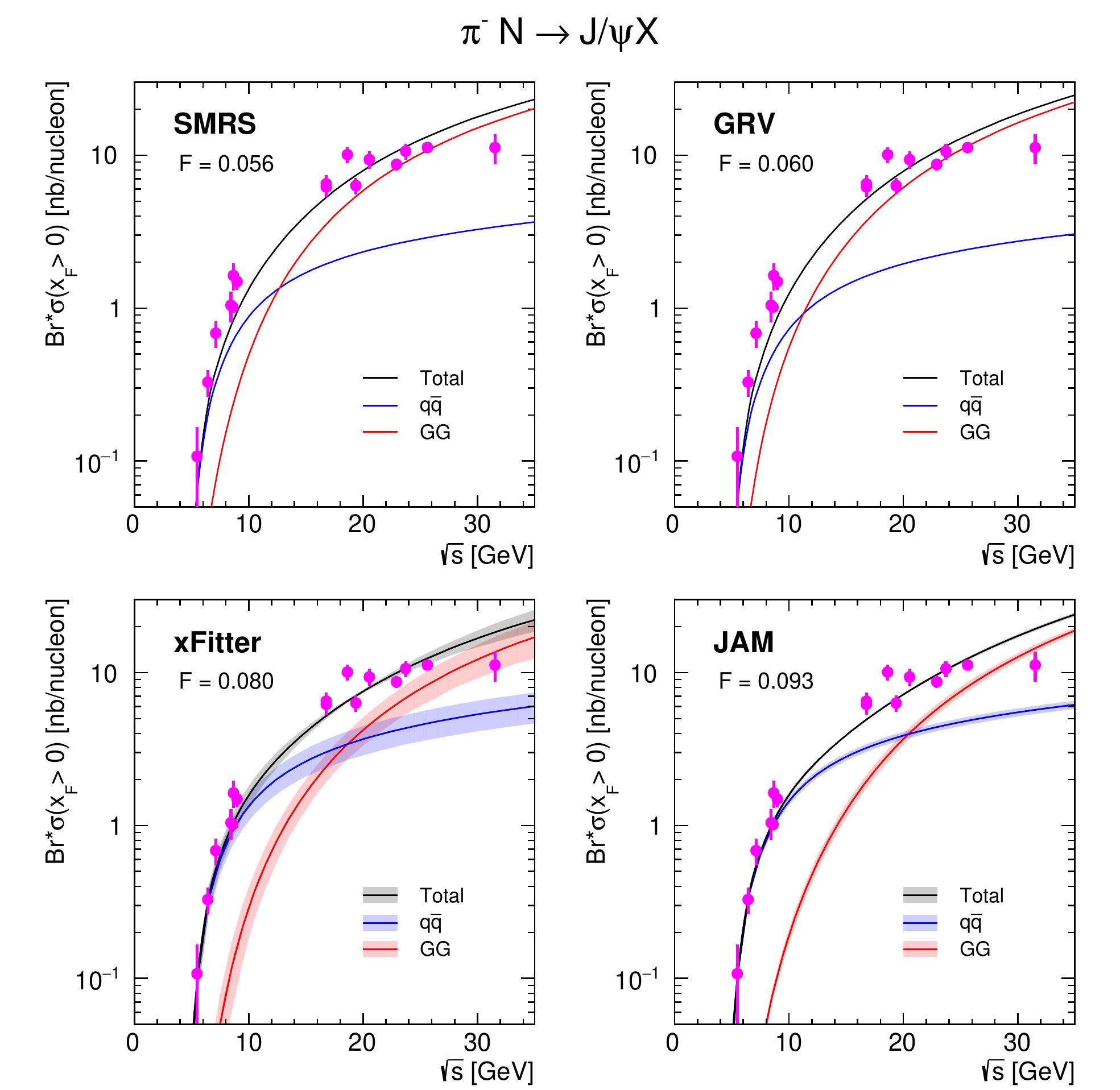}
\caption[\protect{}]{The product of \Jpsi~ dimuon decay branching
  ratio ($\rm{Br}$) and \Jpsi~ production cross sections at $x_F >0$ for
  the $\pi^- N$ reaction, calculated with four pion PDFs (SMRS, GRV,
  xFitter and JAM) is compared with data (solid
  circles)~\cite{Schuler, Alexandrov:1999ch}. The black, blue, and red
  curves represent the calculated total cross section and the $q \bar{q}$
  and $GG$ contributions, respectively. The shaded bands on the
  xFitter and JAM calculations come from the uncertainties of the
  corresponding PDF sets. The SMRS and GRV PDFs contain no information
  on uncertainties.}
\label{fig_jpsi_sdep}
\end{figure}

The results and the comparison with data are displayed in
Fig.~\ref{fig_jpsi_sdep}. The total cross sections for the four PDFs
exhibit quite similar $\sqrt{s}$ dependencies, and all agree
reasonably with the data. The differences between them are visible
through the $F$ factors, which vary from 0.05 to 0.09. As a general
feature, the $q \bar{q}$ contribution dominates at low energies,
whereas the $GG$ contribution becomes increasingly important with
increasing $\sqrt{s}$. However, the relative fractions of $q \bar{q}$
and $GG$ contributions as a function of $\sqrt{s}$ vary considerably,
reflecting the differences between the corresponding parton
distributions. For SMRS and GRV, the $GG$ contribution starts to
dominate the cross section beyond $\sqrt{s}=13$ and $\sqrt{s}=11$ GeV,
respectively, while for xFitter and JAM the corresponding values are
much higher: $\sqrt{s}=19$ and 21 GeV, respectively.

In order to investigate further the effect led by different pion PDFs,
we compare the longitudinal $x_F$ distribution of the calculated
pion-induced \Jpsi~ production cross section with a selection of
fixed-target data from Fermilab and CERN experiments. Among the
datasets available for pion-induced \Jpsi~
production~\cite{jpsi_data1,
  jpsi_data2and3,jpsi_data17and18and21,jpsi_data4and5,jpsi_data17and18and21,
  jpsi_data16, jpsi_data6and7and8, jpsi_data9and10, jpsi_data12and13,
  jpsi_data14and15, jpsi_data19and20}, we choose the ones that have
large-$x_F$ coverage for either hydrogen or light nuclear targets
(lithium and beryllium) in order to minimize the effects of the
nuclear environment. The selected eight datasets are listed in
Table~\ref{tab:data}. The beam momenta of the datasets cover the range
of 39.5--515 GeV/$c$, corresponding to $\sqrt{s}$ values ranging from
8.6 to 31.1 GeV. Some of the data listed in Table~\ref{tab:data}
involve nuclear targets. The target PDFs parametrizations are CT14nlo
for the hydrogen target and EPPS16~\cite{Eskola:2016oht} for the
lithium and beryllium targets. Contrary to the integrated cross
sections, we now allow energy dependence for the hadronization factor
$F$ which is to be fine-tuned for each dataset individually.

\begin{table}[htbp]   
\centering
\begin{tabular}{|c|c|c|c|c|}
\hline
\hline
 Experiment & $P_{beam}$ (GeV/$c$) & Target & Normalization$^{a}$ & References \\
\hline
\hline
FNAL E672, E706 & 515  & Be & 12.0 & \cite{jpsi_data1} \\
FNAL E705  & 300  & Li & 9.5 &\cite{jpsi_data2and3} \\
CERN NA3$^{b}$ & 280  & p  & 13.0 & \cite{jpsi_data17and18and21} \\
CERN NA3$^{b}$  & 200  & p  & 13.0 & \cite{jpsi_data17and18and21} \\
CERN WA11$^{b}$ & 190  & Be  & $^{c}$ & \cite{jpsi_data16} \\
CERN NA3$^{b}$ & 150  & p  & 13.0  & \cite{jpsi_data17and18and21} \\
FNAL E537 & 125  & Be  & 6.0 & \cite{jpsi_data6and7and8} \\
CERN WA39$^{b}$ & 39.5 & p  & 15.0 & \cite{jpsi_data9and10} \\
\hline
\end{tabular}
\caption {The \Jpsi~ production datasets with $\pi^-$ beam used in the analysis, listed in order of decreasing beam momentum.\\
 $^{a}$Percentage of uncertainty in the cross section normalization.\\
 $^{b}$The numerical information was taken from figures. \\
 $^{c}$Information not available.}
\label{tab:data}
\end{table}

Within the CEM and heavy-quark pair production framework introduced in
Sec.~\ref{sec:CEM}, we performed the LO and NLO calculations of the
differential cross sections as a function of $x_F$ with the charm
quark mass $m_c=1.5$ GeV/$c^2$ and renormalization and factorization
scales $\mu_R= m_c$ and $\mu_F = 2 m_c$,
respectively~\cite{Mangano:1992kq}. The comparison of results with the
selected data is shown in
Figs.~\ref{fig_jpsi_data1}-\ref{fig_jpsi_data9}. In
Fig.~\ref{fig_jpsi_data1}, where the dataset has the largest beam
momentum, both LO and NLO CEM results calculated with SMRS, GRV,
xFitter, and JAM pion PDFs are shown, whereas only NLO results are
shown in the other figures.

The hadronization factor $F$, as an overall normalization parameter,
is determined by the best $\chi^2$ fit to the $x_F$ distributions of
cross sections, shown as the black lines in
Figs.~\ref{fig_jpsi_data1}--\ref{fig_jpsi_data9}. The experimental
normalization uncertainties listed in Table~\ref{tab:data} are not
included in the error estimation, since they are correlated systematic
errors and will not affect the $\chi^2$ but only contribute to the
uncertainty of $F$ factor. To compare the four pion PDFs on an equal
footing, the uncertainties of the more recent PDFs are not included in
the calculation of $\chi^2$. We will discuss the impact of PDF
uncertainties on the fit results later.

The $\chi^2$/ndf value of the best fit is also displayed in the
plot. The estimated individual $q \bar{q}$ and $GG$ contributions are
denoted as blue and red lines, respectively. There is a negligible
additional contribution from the $qG$ subprocess, shown as green
lines, to the total cross sections in the NLO calculation. The
calculated value of the $qG$ contribution is
negative~\cite{Nason:1987xz}. The uncertainties of xFitter and JAM PDF
sets are displayed as shaded bands. In the following subsections
(Secs.~\ref{subsec:jpsi_data1}--\ref{subsec:jpsi_data9}), we briefly
comment on the features of each experimental measurement and discuss
the comparison of the data with the CEM calculations. Our observations
are summarized in Sec.~\ref{subsec:observations}.

\subsection{Fermilab E672/E706 experiment}
\label{subsec:jpsi_data1}

The Fermilab E672/E706 experiment~\cite{jpsi_data1} used a 515 GeV/$c$
$\pi^-$ beam scattered off 3.71- and 1.12-cm-long $^{9}$Be
targets. About 9600 \Jpsi~ events integrated in the mass region
between 2.8 and 3.4 GeV/$c^2$ were collected. The final cross sections
cover the range $0.1\leq x_F \leq 0.8$ in bins of 0.02 and have a
normalization uncertainty of 12\%.

\begin{figure}[htbp]
\centering
\includegraphics[width=1.0\columnwidth]{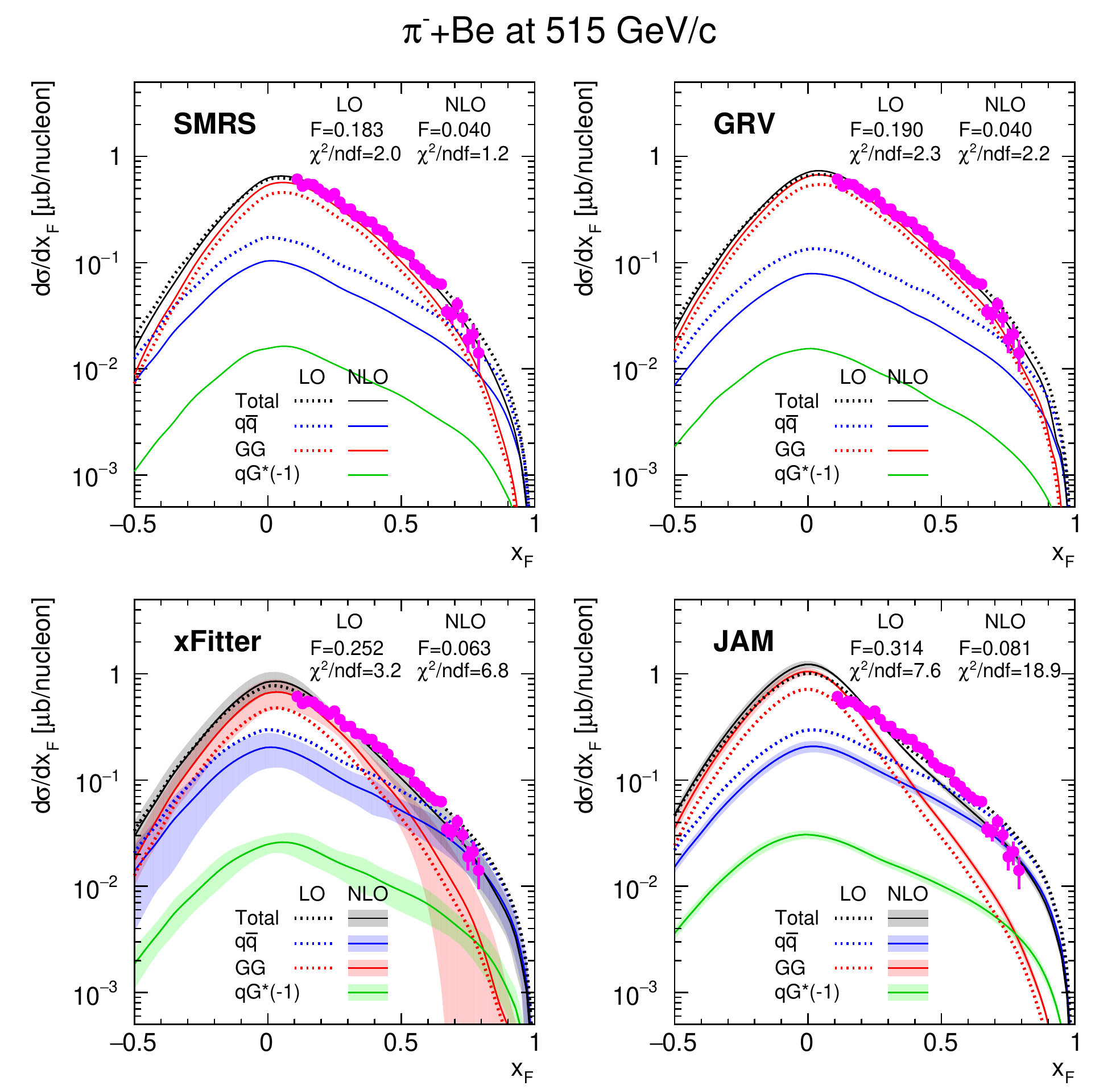}
\caption[\protect{}]{Comparison of the LO and NLO CEM results for the
  SMRS, GRV, xFitter, and JAM PDFs, with the $d\sigma/dx_F$ data of
  \Jpsi~ production off the beryllium target with a 515-GeV/$c$
  $\pi^-$ beam from the E672/E706 experiment~\cite{jpsi_data1}. The
  total cross sections and $q \bar{q}$, $GG$, and $qG\times(-1)$
  contributions are denoted as black, blue, red, and green lines,
  respectively. Solid and dotted lines are for the NLO and LO
  calculations, respectively. The shaded bands on the xFitter and JAM
  calculations come from the uncertainties of the corresponding PDF
  sets. For clarity, the resulting $\chi^2$/ndf and $F$ factors are
  also displayed.}
\label{fig_jpsi_data1}
\end{figure}

The comparison of our calculations at both LO and NLO to the E672/E706
data is shown in Fig.~\ref{fig_jpsi_data1}. Judging from the reduced
$\chi^2$/ndf values, the NLO calculations with SMRS and GRV are in
better agreement with the data than those with xFitter and JAM. The
NLO calculation improves the description of the E672/E706 data only in
the cases of SMRS and GRV. In comparison with the LO, the NLO
calculation has a large effect on the cross sections, increasing its
magnitude by more than a factor of 2. An interesting observation is
that this increase in magnitude is nearly entirely compensated by the
$F$ factor, pointing to a nearly uniform increase along $x_F$. We also
note that the $GG$ contribution dominates the cross section up to
values of $x_F$ as large as 0.5--0.7, depending on the particular pion
PDF set. The additional $qG$ term in the NLO calculation has a minor
(and negative) contribution, although largely dependent on the
particular PDF set.

We observe that the hadronization factor $F$ is reduced by a factor of
4--5 from LO to NLO calculations. Since the NLO calculations involve
higher-order QCD diagrams, the $F$ factor, playing the role of a
normalization constant of cross sections to describe the data, is
expected to be different in the cases of LO and NLO calculations. An
additional justification for the usage of NLO calculations comes from
the fact that all four pion PDFs examined in this work are determined
in a NLO global analysis.

\subsection{Fermilab E705 experiment}
\label{subsec:jpsi_data2}

The Fermilab E705 experiment~\cite{jpsi_data2and3} used a 300 GeV/$c$
negative hadron beam (with 98\% pions) scattered off a 33-cm-long
lithium target. Data were also collected with a positive hadron beam
consisting of protons and positive pions. Thanks to the open geometry
spectrometer, an excellent mass resolution was achieved, allowing a
measurement of the \Jpsi~ peak in the mass range between 2.98 and 3.18
GeV/$c^2$. Since the final number of \Jpsi~ events was not explicitly
given, we estimate it from the published statistical errors to about
6000 events for the negative pion beam. The final cross sections have
a normalization uncertainty of 11.1\% and cover the range $-0.1\leq
x_F \leq 0.45$ in bins of 0.05.

\begin{figure}[htbp]
\includegraphics[width=1.0\columnwidth]{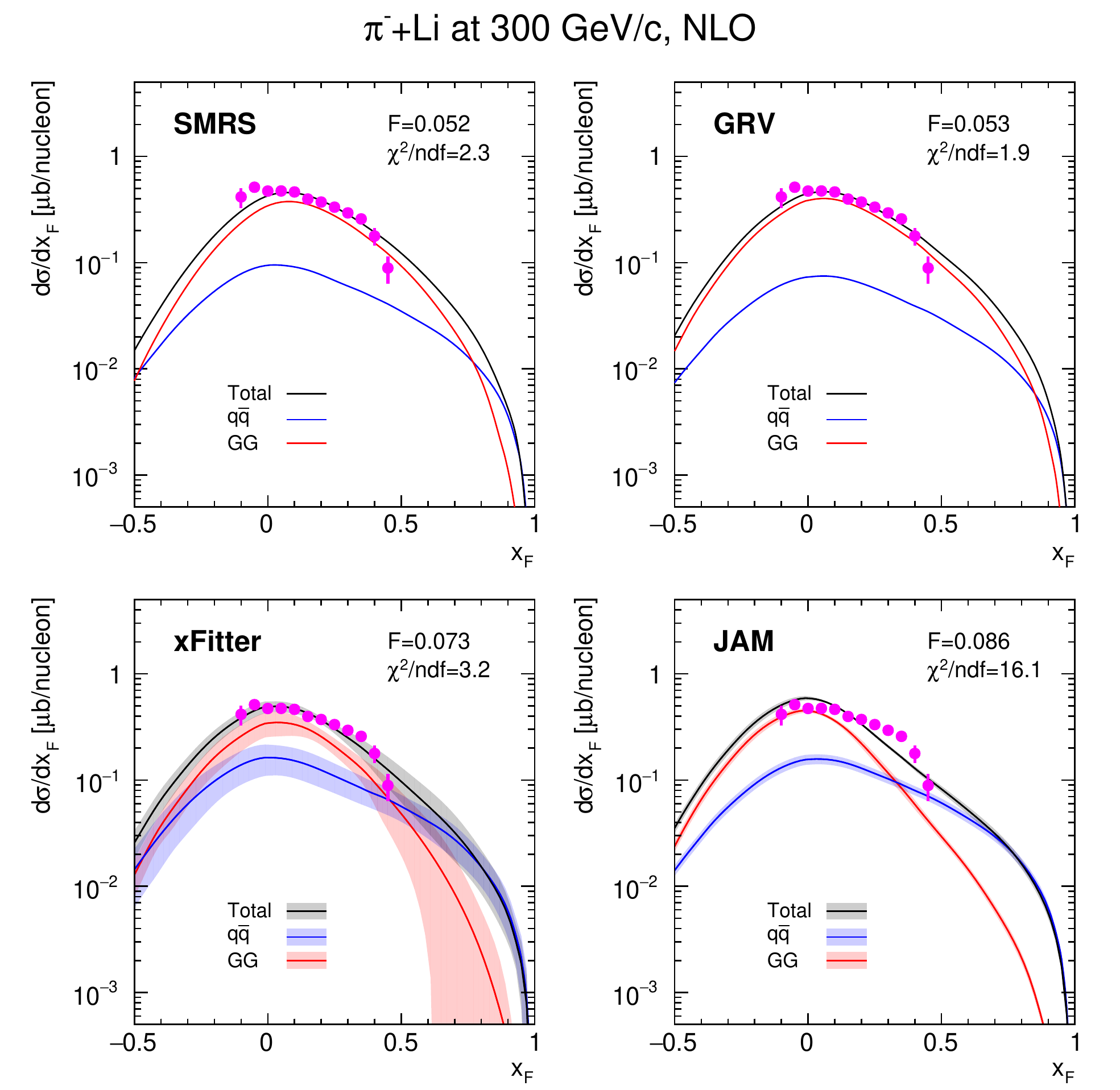}
\caption
[\protect{}]{Comparison of the NLO CEM results for the SMRS, GRV,
  xFitter, and JAM PDFs with $d\sigma/dx_F$ data of \Jpsi~ production
  off the lithium target with a 300-GeV/$c$ $\pi^-$ beam from the E705
  experiment~\cite{jpsi_data2and3}. The total cross sections and $q
  \bar{q}$ and $GG$ contributions are denoted as black, blue, and red
  lines, respectively.}
\label{fig_jpsi_data2}
\end{figure}

The comparison of our calculations with the experimental cross
sections is shown in Fig.~\ref{fig_jpsi_data2}. The best $\chi^2$/ndf
value is obtained with the SMRS PDFs. In contrast, the use of the JAM
PDFs results in a significantly degraded $\chi^2$/ndf. The $GG$
contribution for the JAM PDFs has a falloff in $x_F$ too fast to
describe the data. We observe a trend similar to the one seen already
in Fig.~\ref{fig_jpsi_data1}: The crossover between the central values
of $q \bar{q}$ and $GG$ terms for SMRS and GRV occurs at values of
$x_F$ much larger than the ones for xFitter and JAM.

\subsection{CERN NA3 experiment, 280 GeV/$c$}
\label{subsec:jpsi_data21}

The CERN NA3 experiment~\cite{jpsi_data17and18and21}, performed nearly
four decades ago, still has the largest pion-induced \Jpsi~ production
statistics available today. Data were taken at three different
incident momenta, 280, 200, and 150 GeV/$c$ with both positive and
negative hadron beams. The beam components were identified using
Cherenkov counters. Moreover, in addition to a heavy platinum target,
a liquid hydrogen target was also used, thus eliminating all possible
nuclear effects. For all three energies, the cross sections have a
normalization uncertainty of 13\%. In the present study we consider
only the NA3 hydrogen data. Unfortunately, these invaluable numerical
cross sections were never published and could be retrieved only from
the figures in the published paper~\cite{jpsi_data17and18and21} and
unpublished thesis~\cite{PhD-PhC}.

\begin{figure}[htbp]
\includegraphics[width=1.0\columnwidth]{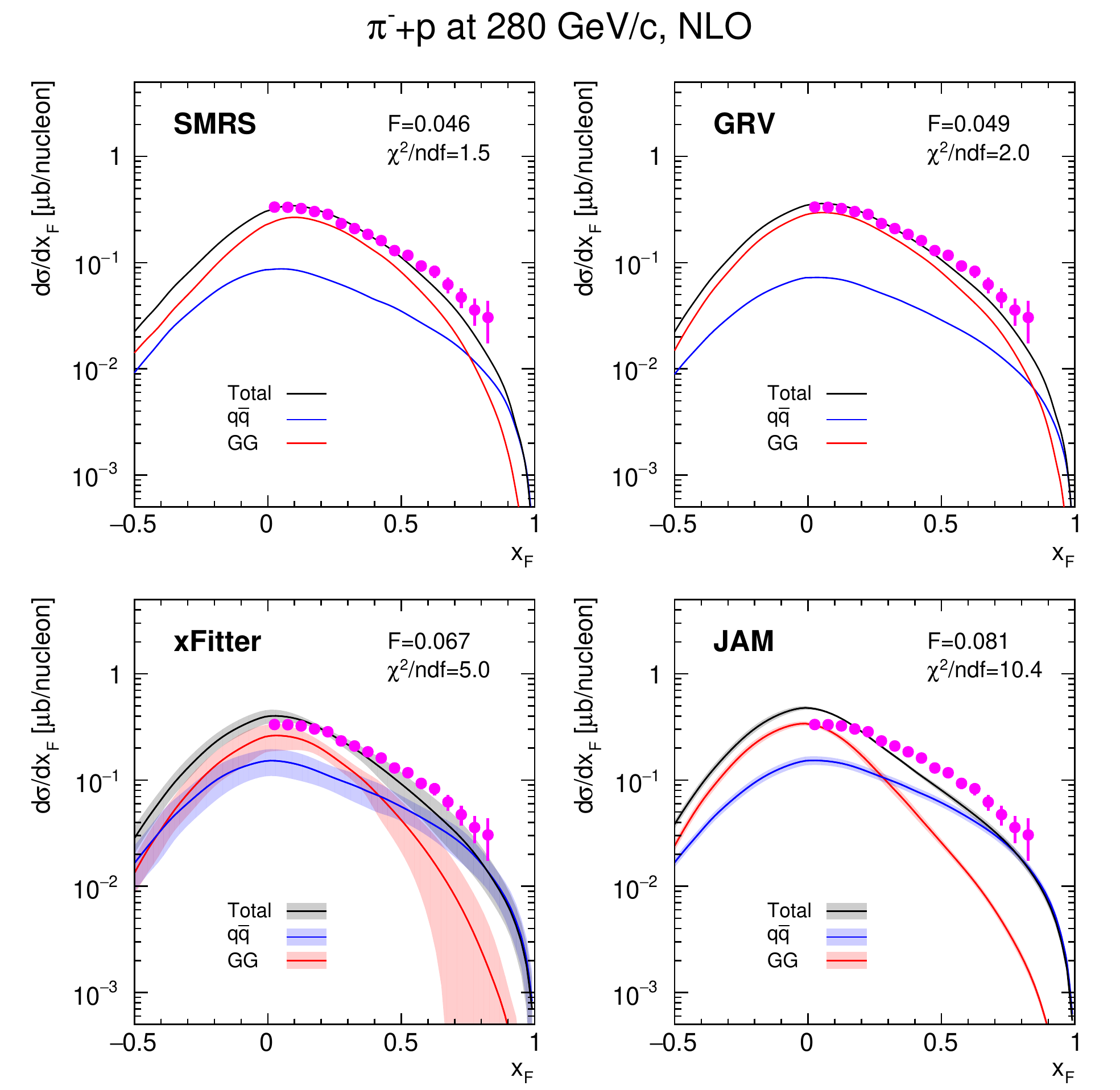}
\caption[\protect{}]{Comparison of the NLO CEM results for the SMRS,
  GRV, xFitter and JAM PDFs, with the $d\sigma/dx_F$ data of \Jpsi~
  production off the hydrogen target with a 280-GeV/$c$ $\pi^-$ beam
  from the NA3 experiment~\cite{jpsi_data17and18and21}. The total
  cross sections and $q \bar{q}$ and $GG$ contributions are denoted as
  black, blue, and red lines, respectively.}
\label{fig_jpsi_data21}
\end{figure}

For the 280 GeV/$c$ data taking, the authors used a 50-cm-long
hydrogen target, resulting in 23350 \Jpsi~ $\pi^-$ events in the
dimuon mass region between 2.7 and 3.5 GeV/$c^2$. The retrieved data
are available in 17 $x_F$ bins of 0.05, between 0.025 and 0.825. The
comparison with the NLO CEM calculation is shown in
Fig.~\ref{fig_jpsi_data21}. The resulting $\chi^2$/ndf values repeat
the trend already observed: They are better for the calculations with
SMRS and GRV PDFs and are 2--4 times larger for xFitter and JAM. We
note in passing that the relatively small $\chi^2$/ndf values could
partly be caused by the overestimation of the statistical errors in
retrieving the original cross sections from the published figures.

\subsection{CERN NA3 experiment, 200 GeV/$c$}
\label{subsec:jpsi_data17}

The data at 200 GeV/$c$ incident momentum were taken with a 30-cm-long
hydrogen target. With the negative hadron beam 3157 pion-induced
\Jpsi~ events were collected. The retrieved data extend from $x_F =
0.05$ to $x_F = 0.85$.

\begin{figure}[htbp]
\includegraphics[width=1.0\columnwidth]{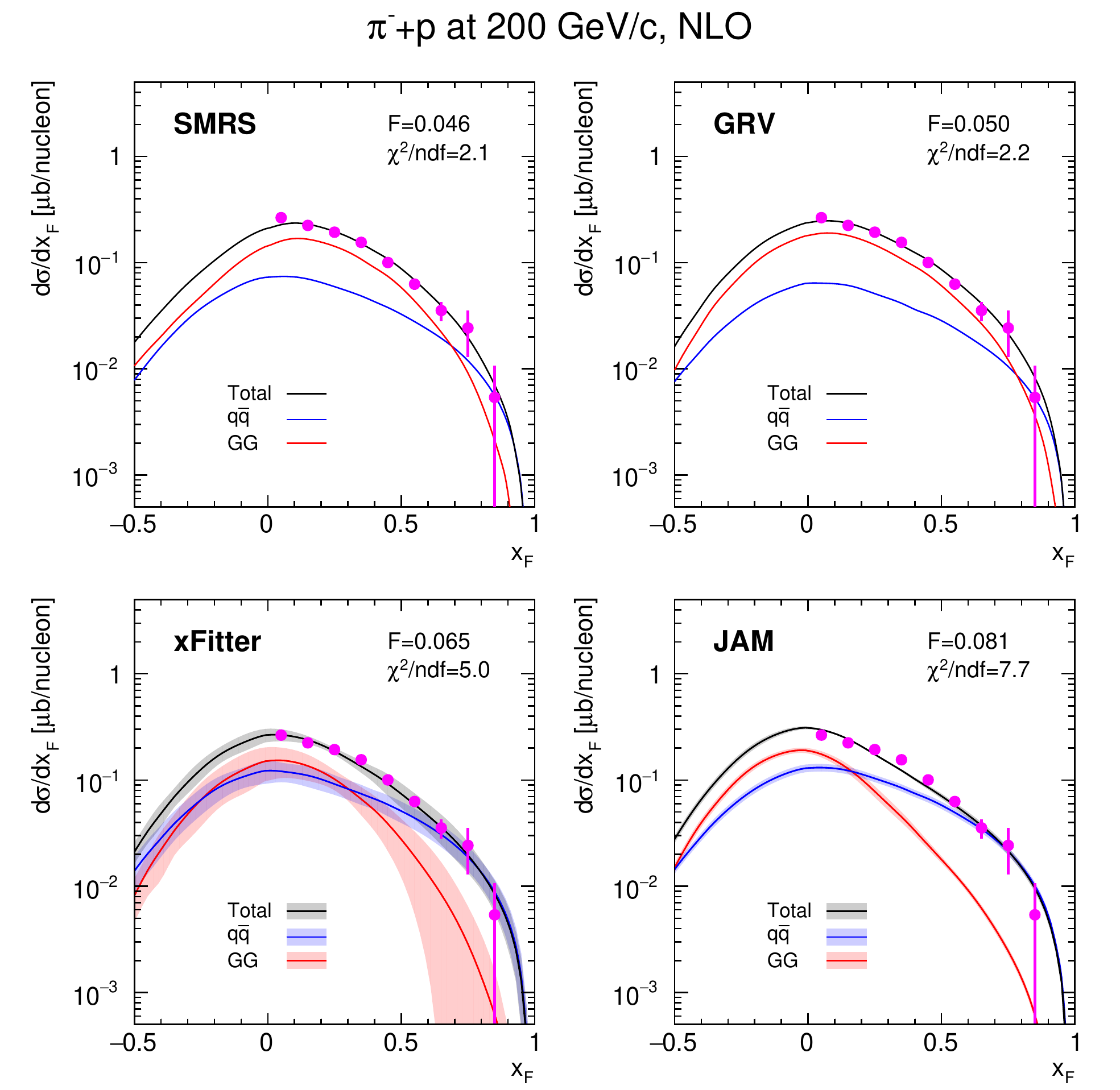}
\caption[\protect{}]{Comparison of the NLO CEM results for the SMRS,
  GRV, xFitter, and JAM PDFs, with the $d\sigma/dx_F$ data of \Jpsi~
  production off the hydrogen target with a 200-GeV/$c$ $\pi^-$ beam
  from the NA3 experiment~\cite{jpsi_data17and18and21}. The total
  cross sections and $q \bar{q}$ and $GG$ contributions are denoted as
  black, blue, and red lines, respectively.}
\label{fig_jpsi_data17}
\end{figure}

The comparison of the NLO calculation with the data is shown in
Fig.~\ref{fig_jpsi_data17}. The agreement with the data is fair for
all PDF sets, although the general trend persists: The most recent
xFitter and JAM global fits have slightly worse $\chi^2$/ndf
values. We also note that, as the incident momentum decreases, the
importance of the $q \bar{q}$ term increases, particularly for the
larger values of $x_F$. The $GG$ contribution dominates the cross
section for the calculation with the GRV PDFs up to $x_F=0.6$. In
contrast, for the JAM PDFs, the corresponding value is much lower:
$x_F=0.2$.

\subsection{CERN WA11 experiment}
\label{subsec:jpsi_data16}

The WA11 Collaboration at CERN measured \Jpsi~ production cross
sections~\cite{jpsi_data16} using a 190 GeV/$c$ negative pion beam
scattered off a triplet of beryllium target with a total length of 8.9
cm. Thanks to the open spectrometer geometry used, an excellent \Jpsi~
mass resolution, $\sigma=31$ MeV/$c^2$, was achieved. The large
spectrometer coverage in dimuon opening angles made possible
measurements at $x_F$ values from --0.35 to 0.75, in bins of
0.10. About 38000 \Jpsi~ events were reported in the mass range
between 3.00 and 3.18 GeV/$c^2$, including 7\% background. The same
experiment had previously measured the feed-down contribution from the
$\chi_c$ decays. In the cross sections shown, this contribution was
subtracted. For consistency, the reported feed-down contributions were
added to the prompt cross section values shown, using the described
procedure in reverse order.

\begin{figure}[htbp]
\includegraphics[width=1.0\columnwidth]{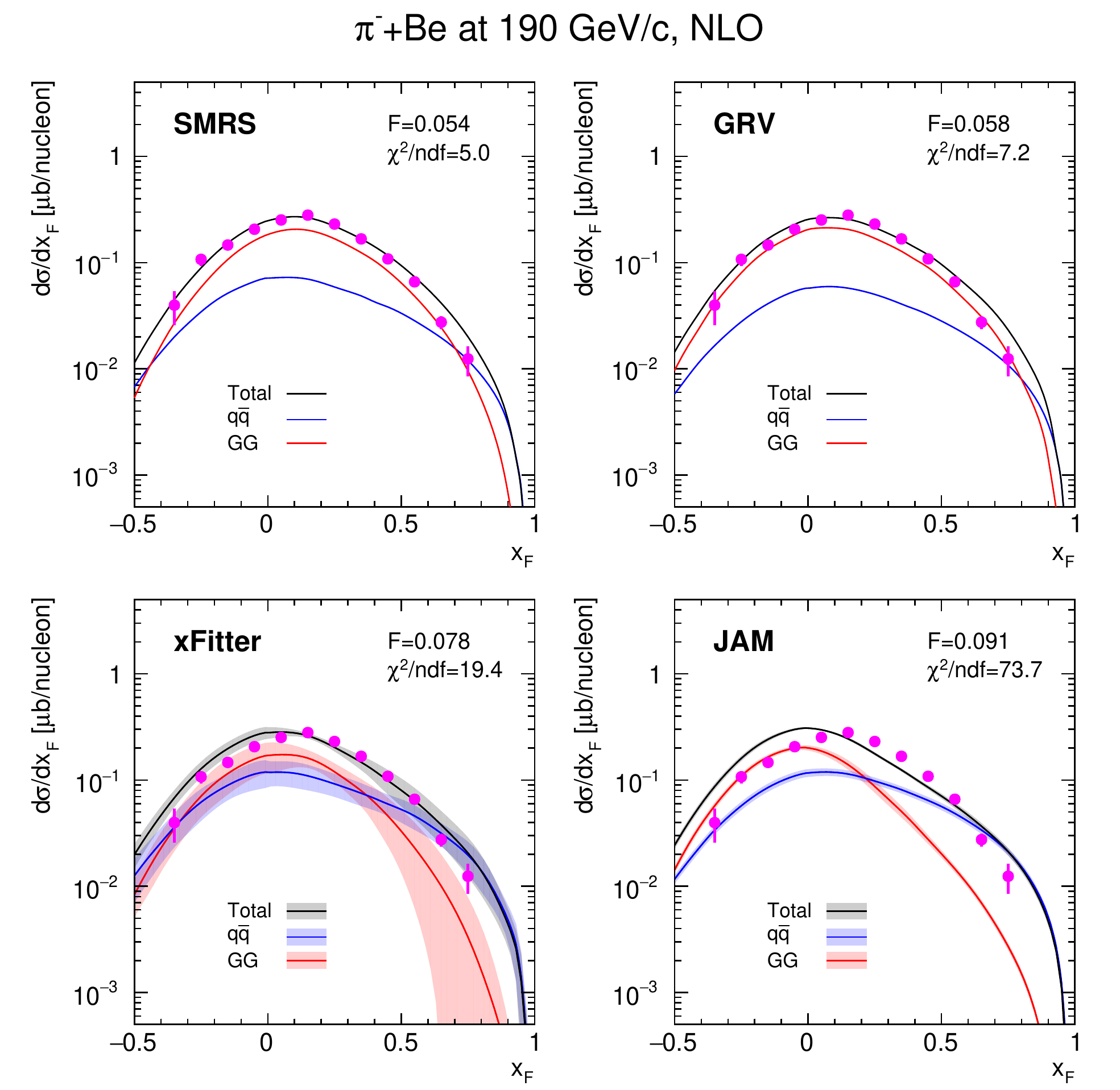}
\caption[\protect{}]{Comparison of the NLO CEM results for the SMRS,
  GRV, xFitter, and JAM PDFs, with the $d \sigma / dx_F$ data of
  \Jpsi~ production off the beryllium target with a 190-GeV/$c$
  $\pi^-$ beam from the WA11 experiment~\cite{jpsi_data16}. The total
  cross sections and $q \bar{q}$ and $GG$ contributions are denoted as
  black, blue, and red lines, respectively.}
\label{fig_jpsi_data16}
\end{figure}

The comparison of the NLO CEM calculations with the WA11 data is shown
in Fig.~\ref{fig_jpsi_data16}. The resulting $\chi^2$/ndf values are
larger than for the NA3 data, pointing to additional systematic errors
either in the original data or in the procedure of retrieving
them. Not surprisingly, however, the overall conclusions are similar
to the ones made previously for the 200 GeV/$c$ data. The calculations
with SMRS and GRV are in better agreement with the data than xFitter
and JAM.

\subsection{CERN NA3 experiment, 150 GeV/$c$}
\label{subsec:jpsi_data26}

The NA3 data at 150 GeV/$c$ were taken with a 30-cm-long hydrogen
target. The statistics is large, as 16952 events were reported. The
original data cover the $x_F$ region between 0.025 and 0.975, in bins
of 0.05. The data retrieved from the published figures extend to
$x_F=0.925$.

\begin{figure}[htbp]
\includegraphics[width=1.0\columnwidth]{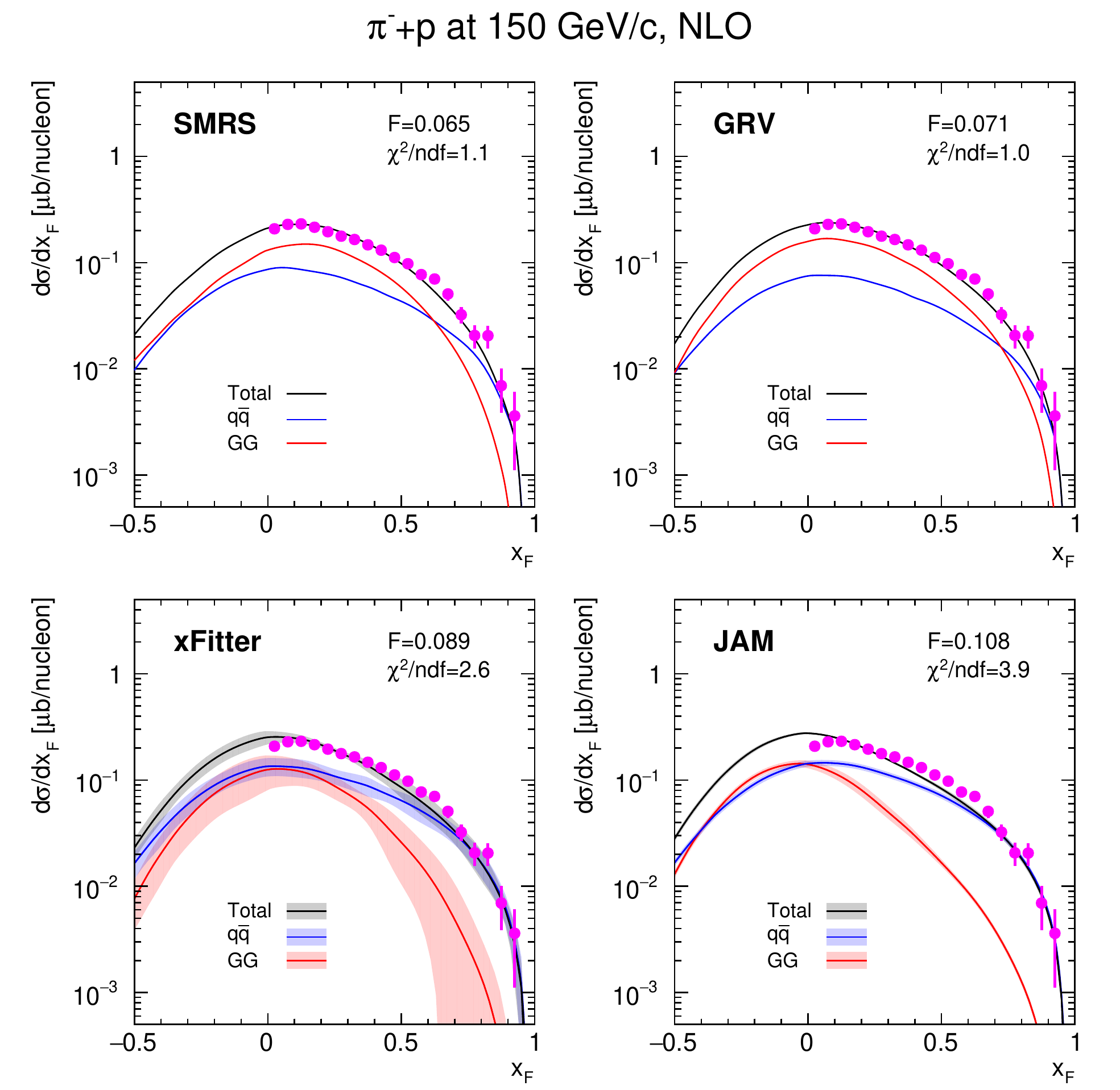}
\caption[\protect{}]{Comparison of the NLO CEM results for the SMRS,
  GRV, xFitter and JAM PDFs, with the $d\sigma/dx_F$ data of \Jpsi~
  production off the hydrogen target with a 150-GeV/$c$ $\pi^-$ beam
  from the NA3 experiment~\cite{jpsi_data17and18and21}. The total
  cross sections and $q \bar{q}$ and $GG$ contributions are denoted as
  black, blue, and red lines, respectively.}
\label{fig_jpsi_data26}
\end{figure}

The comparison with the NLO CEM calculation is shown in
Fig.~\ref{fig_jpsi_data26}. The calculated $\chi^2$/ndf values are
rather small, pointing to somewhat overestimated experimental error
bars. Nevertheless, they remain larger for the two most recent PDF
sets. The overall trend previously observed is confirmed.

\subsection{Fermilab E537 experiment}
\label{subsec:jpsi_data8}

The E537 experiment at Fermilab has measured \Jpsi~ production cross
sections induced by a hadron beam of 125 GeV/$c$ containing 82\%
negative pions and 18\% antiprotons. Three different targets have been
used: beryllium, copper, and tungsten. An experimental mass resolution
of $\sigma=200$ MeV/$c^2$ for the Be target is reported. The 2881
collected events with the Be target in the region of the \Jpsi~ peak
cover the $x_F$ region between 0.05 and 0.95, in bins of 0.10. The
normalization uncertainty on the cross sections is 6\%.

\begin{figure}[htbp]
\includegraphics[width=1.0\columnwidth]{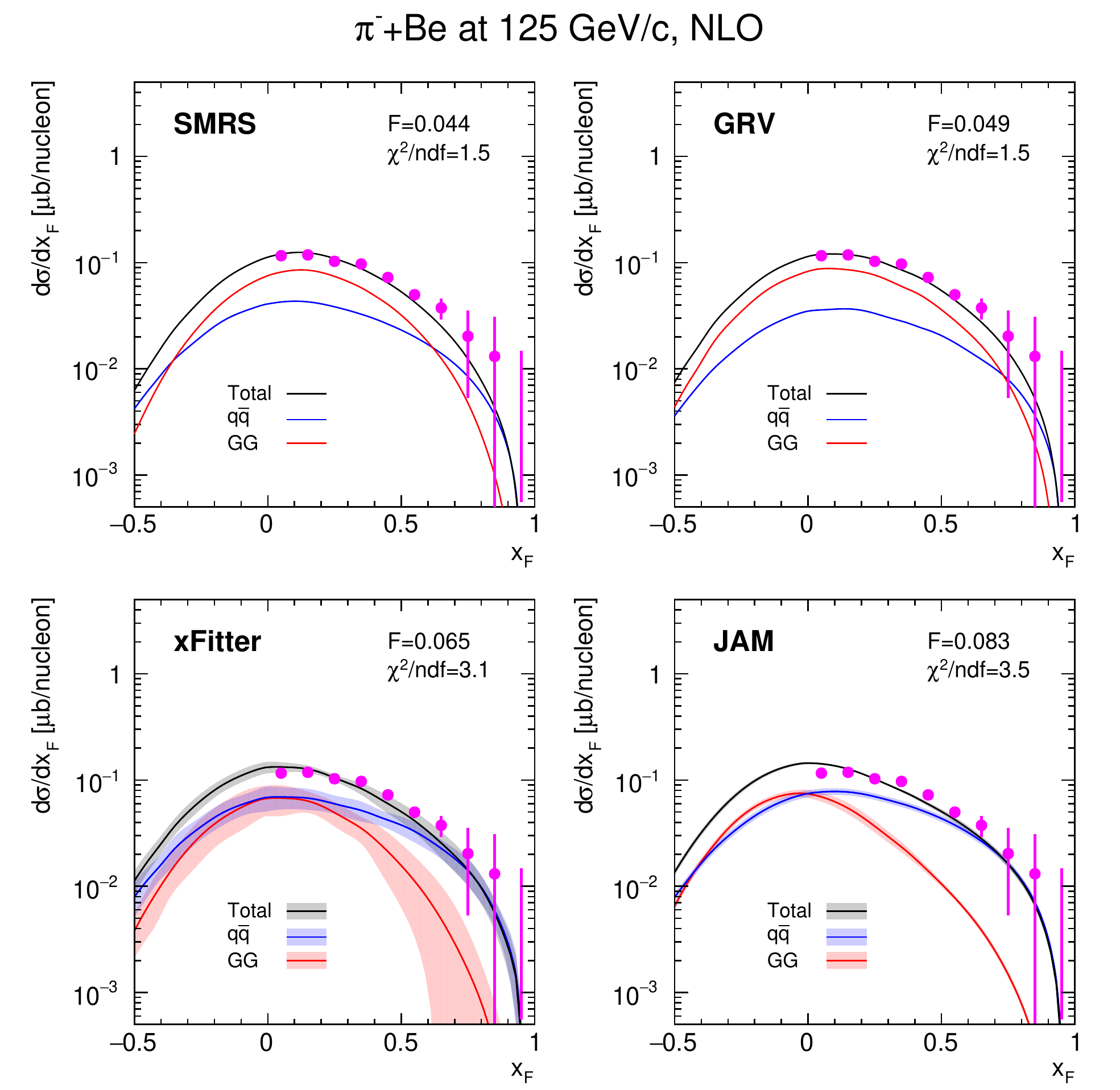}
\caption[\protect{}]{Comparison of the NLO CEM results for the SMRS,
  GRV, xFitter, and JAM PDFs, with the $d\sigma/dx_F$ data of \Jpsi~
  production off the hydrogen target with a 125-GeV/$c$ $\pi^-$ beam
  from the E537 experiment~\cite{jpsi_data6and7and8}. The total cross
  sections and $q \bar{q}$ and $GG$ contributions are denoted as
  black, blue, and red lines, respectively.}
\label{fig_jpsi_data8}
\end{figure}

The NLO CEM calculation and the E537 data are shown in
Fig.~\ref{fig_jpsi_data8}. The $\chi^2$/ndf values are reasonable for
SMRS and GRV calculations and again slightly worse for xFitter and
JAM. For values of $x_F \simeq 0$, the magnitude of the $q \bar{q}$
term is similar to that of the $GG$ term. We also observe the
relatively quick decrease of the $GG$ term for the calculation with
the JAM gluon PDF.

\subsection{CERN WA39 experiment}
\label{subsec:jpsi_data9}

The CERN WA39 Collaboration measured the \Jpsi~ production cross
section with a 39.5 GeV/$c$ hadron beam momentum. Data for the
67-cm-long liquid hydrogen target were taken with negative and
positive hadron beams. Measurements are reported with incident
$\pi^+$, $\pi^-$, K$^+$, K$^-$, $p$ and $\bar p$. Most of the 402 events
reported for the negative hadron beam are pion-induced \Jpsi's. The
$x_F$-differential cross sections, available as a figure in the
published paper, cover the region $0.05 \leq x_F \leq 0.85$ in bins of
0.10. The normalization uncertainty on the cross sections is 15\%.

\begin{figure}[htbp]
\includegraphics[width=1.0\columnwidth]{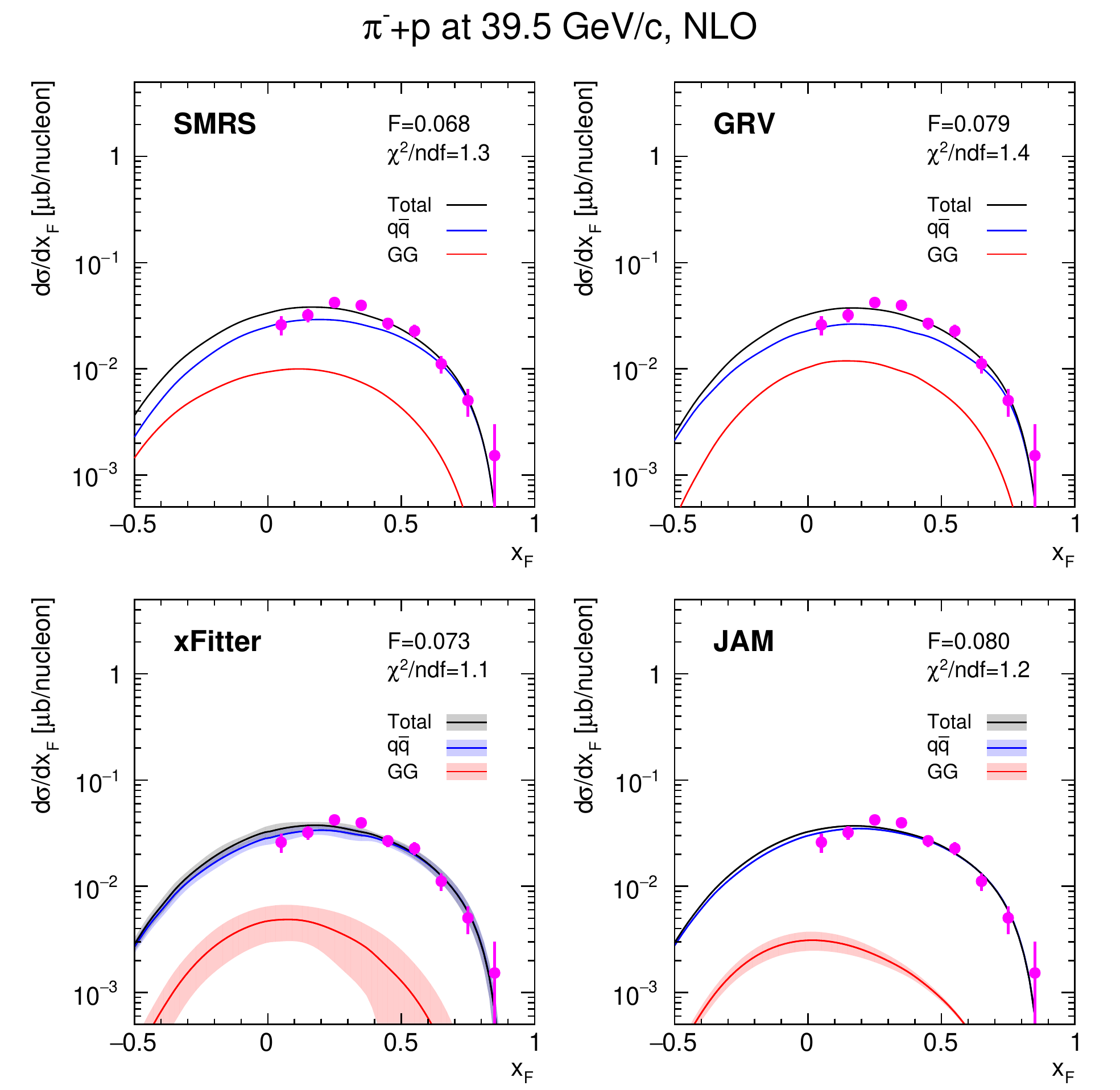}
\caption[\protect{}]{Comparison of the NLO CEM results for the SMRS,
  GRV, xFitter, and JAM PDFs, with the $d\sigma/dx_F$ data of \Jpsi~
  production off the hydrogen target with a 39.5-GeV/$c$ $\pi^-$ beam
  from the WA39 experiment~\cite{jpsi_data9and10}. The total cross
  sections and $q \bar{q}$ and $GG$ contributions are denoted as
  black, blue and red lines, respectively.}
\label{fig_jpsi_data9}
\end{figure}

The comparison between data and calculations is shown in
Fig.~\ref{fig_jpsi_data9}. The immediate observation is that for this
low incident momentum the $q \bar{q}$ contribution is much larger than
the $GG$ term, by a factor of 5--8 around $x_F = 0$. The $\chi^2$/ndf
values for the four PDFs are all close to 1 and slightly larger for
the calculation with SMRS.

\subsection{Observations}
\label{subsec:observations}

As a general observation, both LO and NLO CEM calculations provide a
reasonable description of $x_F$ distributions of \Jpsi~ production in
the energy range considered
(Figs.~\ref{fig_jpsi_data1}--\ref{fig_jpsi_data9}). We note that the
large difference in the magnitude between LO and NLO is compensated by
the $F$ factor. The $F$ factors for the xFitter and JAM PDFs are
relatively stable across the range of collision energies, while the
factors for SMRS and GRV PDFs show a mild rise toward low
energies. From the comparison between data and calculations,
interesting observations are summarized below.

(i) The importance of the $GG$ contribution relative to that of $q
\bar{q}$ is greatly enhanced in the NLO calculation. As for the
description of the large-$x_F$ data points for the pion beam larger
than 125 GeV/$c$, the $\chi^2$/ndf values with the NLO calculations
generally improve for the results with SMRS and GRV, whereas those
with xFitter and JAM become worse, compared to the LO ones; for
example, see Fig.~\ref{fig_jpsi_data1}.

(ii) At low energies, the $GG$ contribution is relatively small, but
it increases rapidly with the increase of energy. The fraction of $GG$
component is maximized around $x_F = 0$, corresponding to the gluon
density of pions at $x \sim$0.1--0.2. As a result of the rapid drop of
the pion's gluon density toward $x=1$ shown in Fig.~\ref{fig_pdf}(c),
the $GG$ contribution decreases dramatically toward large $x_F$. In
contrast, the $q \bar{q}$ contribution falls off slower at high $x_F$
because of the relatively strong pion valence antiquark density at
large $x$. Consequently, the $q \bar{q}$ contribution has a broader
$x_F$ distribution than that of the $GG$ contribution and the relative
importance of $q \bar{q}$ rises at the large-$x_F$ region. As
mentioned before, CEM dictates the relative weighting between $q
\bar{q}$ and $GG$ subprocesses by the convolution of pQCD calculation
and parton densities, and the $F$ factor cannot modify the shape of
$d\sigma/d x_F$. Therefore, adequate shapes of $d\sigma/d x_F$
distributions of individual $GG$ and $q \bar{q}$ contributions from
CEM calculations are required to achieve a reasonable description of
data points at $x_F>0.5$. Since the partonic cross sections and
nucleon PDFs are basically common in the calculations
[Eq.~(\ref{eq:eq1})], the variation of results shall originate from
the difference in the folded pion partonic densities. The calculations
with SMRS and GRV pion PDFs agree with the data overall, while
significantly large $\chi^2$/ndf values are found in the description
of data with a beam momentum greater than 125 GeV/$c$ for both xFitter
and JAM pion PDFs.

(iii) At low beam energies such as 39.5 GeV/$c$ in Fig.~\ref{fig_jpsi_data9},
the $q\bar{q}$ process is the dominant mechanism of \Jpsi~ production
over the whole $x_F$ region. The data are much less, if at all,
sensitive to the variation of the $GG$ contribution. Good $\chi^2$/ndf
values are obtained for all four pion PDFs.

\section{Systematic Study}
\label{sec:syst}

Through the comparison of data with calculations over a broad energy
range, we have two major findings: (i) The large-$x_F$ distribution of
\Jpsi~ production is sensitive to the pion gluon density; (ii) the
central values of gluon densities of the recently available JAM and
xFitter fall off too rapidly at large $x$ and fail to describe the
$x_F$ distributions of \Jpsi~ data. Judging from the consistency of
observation for the datasets with proton and nuclear targets, the
unaccounted nuclear medium effects such as the energy loss effect are
unlikely to change the conclusions.

The uncertainties provided by xFitter and JAM can have an impact on
the $\chi^2$ values of the fits. We perform a new fit where the PDF
uncertainties are added as theoretical errors into the covariance
matrix of measurements for the calculation of the $\chi^2$
values. Table~\ref{tab:pdf_uncertainties} lists the best-fit $F$
factor and $\chi^2$/ndf values for the four pion PDFs without PDF
uncertainties, as shown in
Figs.~\ref{fig_jpsi_data1}--\ref{fig_jpsi_data9}, and those with
inclusion of PDF uncertainties for xFitter and JAM. As expected, the
$\chi^2$/ndf value of the fit for both xFitter and JAM improves after
taking into account the PDF uncertainties. The $F$ factor remains
basically unchanged. Because of the relatively large uncertainties
assigned by xFitter, the improvement of $\chi^2$/ndf is more
pronounced for xFitter than JAM, even though these two pion PDFs have
similar central values in their large-$x$ gluon distributions.

\begin{table}[htbp]   
\centering
\begin{tabular}{|c|c|c|c|c|c|c|c|c|c|c|c|c|}
\hline
Data  & \multicolumn{2}{|c|}{SMRS} & \multicolumn{2}{|c|}{GRV} & \multicolumn{4}{|c|}{xFitter} & \multicolumn{4}{|c|}{JAM} \\
\hline
Experiment ($P_{beam}$) & $F$  & $\chi^2$/ndf & $F$  & $\chi^2$/ndf & $F$ & $F$$^{*}$  & $\chi^2$/ndf & $\chi^2$/ndf$^{*}$ & $F$ & $F$$^{*}$  & $\chi^2$/ndf & $\chi^2$/ndf$^{*}$ \\
\hline
E672, E706 (515) &  0.040 &  1.2 & 0.040 &  2.2 & 0.063 & 0.063 &  6.8 &  4.7 & 0.081 & 0.081 & 18.9 & 18.5 \\
E705 (300) &  0.052 &  2.3 & 0.053 &  1.9 & 0.073 & 0.076 &  3.2 &  1.3 & 0.086 & 0.086 & 16.1 & 15.9 \\
NA3 (280) &   0.046 &  1.5 & 0.049 &  2.0 & 0.067 & 0.069 &  5.0 &  3.2 & 0.081 & 0.081 & 10.4 & 10.3 \\
NA3 (200) &   0.046 &  2.1 & 0.050 &  2.2 & 0.065 & 0.066 &  5.0 &  1.3 & 0.081 & 0.081 &  7.7 &  7.6 \\
WA11 (190) &  0.054 &  5.0 & 0.058 &  7.2 & 0.078 & 0.076 & 19.4 &  6.2 & 0.091 & 0.091 & 73.7 & 72.9 \\
NA3 (150) &   0.065 &  1.1 & 0.071 &  1.0 & 0.089 & 0.091 &  2.6 &  1.6 & 0.108 & 0.108 &  3.9 &  3.8 \\
E537 (125) &  0.044 &  1.5 & 0.049 &  1.5 & 0.065 & 0.065 &  3.1 &  1.4 & 0.083 & 0.083 &  3.5 &  3.5 \\
WA39 (39.5) &  0.068 &  1.3 & 0.079 &  1.4 & 0.073 & 0.072 &  1.1 &  0.8 & 0.080 & 0.080 &  1.2 &  1.2 \\
\hline
\end{tabular}
\caption {Results of $F$ factor and $\chi^2$/ndf value of the best fit
  of the NLO CEM calculations for SMRS, GRV, xFitter, and JAM pion
  PDFs to the data listed in Table~\ref{tab:data}. The $F$$^{*}$
  factor and $\chi^2$/ndf$^{*}$ are the ones corresponding to the fit
  with inclusion of PDF uncertainties for xFitter and JAM.}
\label{tab:pdf_uncertainties}
\end{table}

To check the sensitivity of the CEM calculation to various QCD
parameters and the choice of nuclear PDFs, we have performed a
systematic study. Taking the convention of the charm quark mass in
Refs.~\cite{Gavai:1994in, Schuler:1996ku, Nelson:2012bc}, we test the
variations of results by setting $m_c$ to be 1.2 GeV/$c^2$. The
dependence on the renormalization scale $\mu_R$ is checked by varying
at 0.5, 1.0, and 2.0 $m_c$~\cite{Mangano:1992kq}. We also make a
different choice of nCTEQ15~\cite{Kovarik:2015cma} as the nuclear PDF
in the calculations with GRV, JAM, and xFitter. Overall, the above
observations remain qualitatively valid with respect to all these
systematic variations.

\begin{figure}[htbp]
\includegraphics[width=0.75\columnwidth]{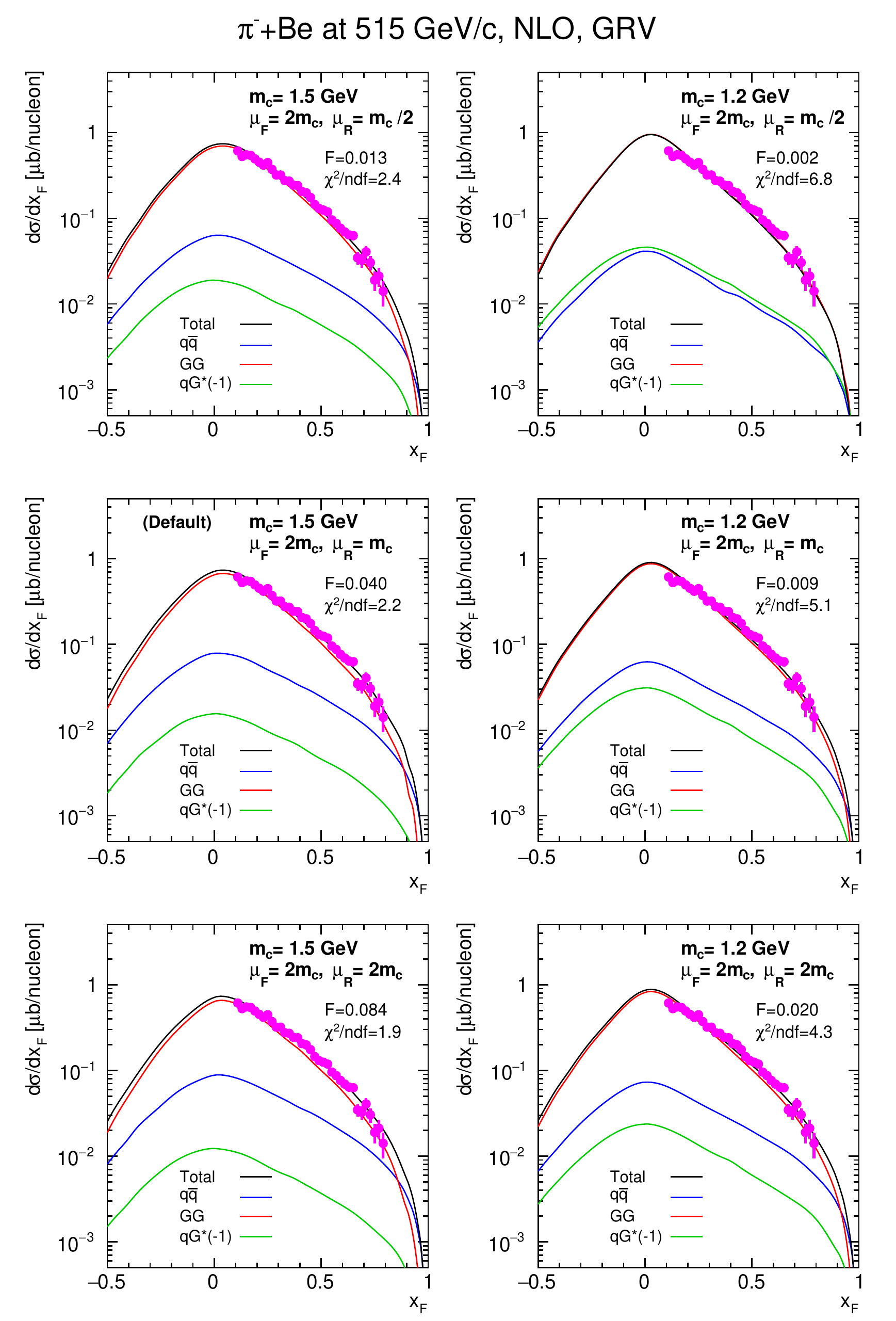}
\caption[\protect{}]{The NLO CEM results with variation of charm quark
  mass $m_c$ and renormalization scale $\mu_R$, compared with the
  $d\sigma/dx_F$ data of \Jpsi~ production off the beryllium target
  with a 515-GeV/$c$ $\pi^-$ beam from the E672/E706
  experiment~\cite{jpsi_data1}. The pion PDFs used for the calculation
  is GRV. The total cross sections and $q \bar{q}$, $GG$, and
  $qG\times(-1)$ contributions are denoted as black, blue, red and
  green lines, respectively. The charm quark mass $m_c$, factorization
  scale $\mu_F$, and renormalization scale $\mu_R$ used for the CEM
  calculation as well as the fit $\chi^2$/ndf and $F$ factors are
  displayed in each plot.}
\label{fig3_sys_GRV}
\end{figure}

\begin{figure}[htbp]
\includegraphics[width=0.75\columnwidth]{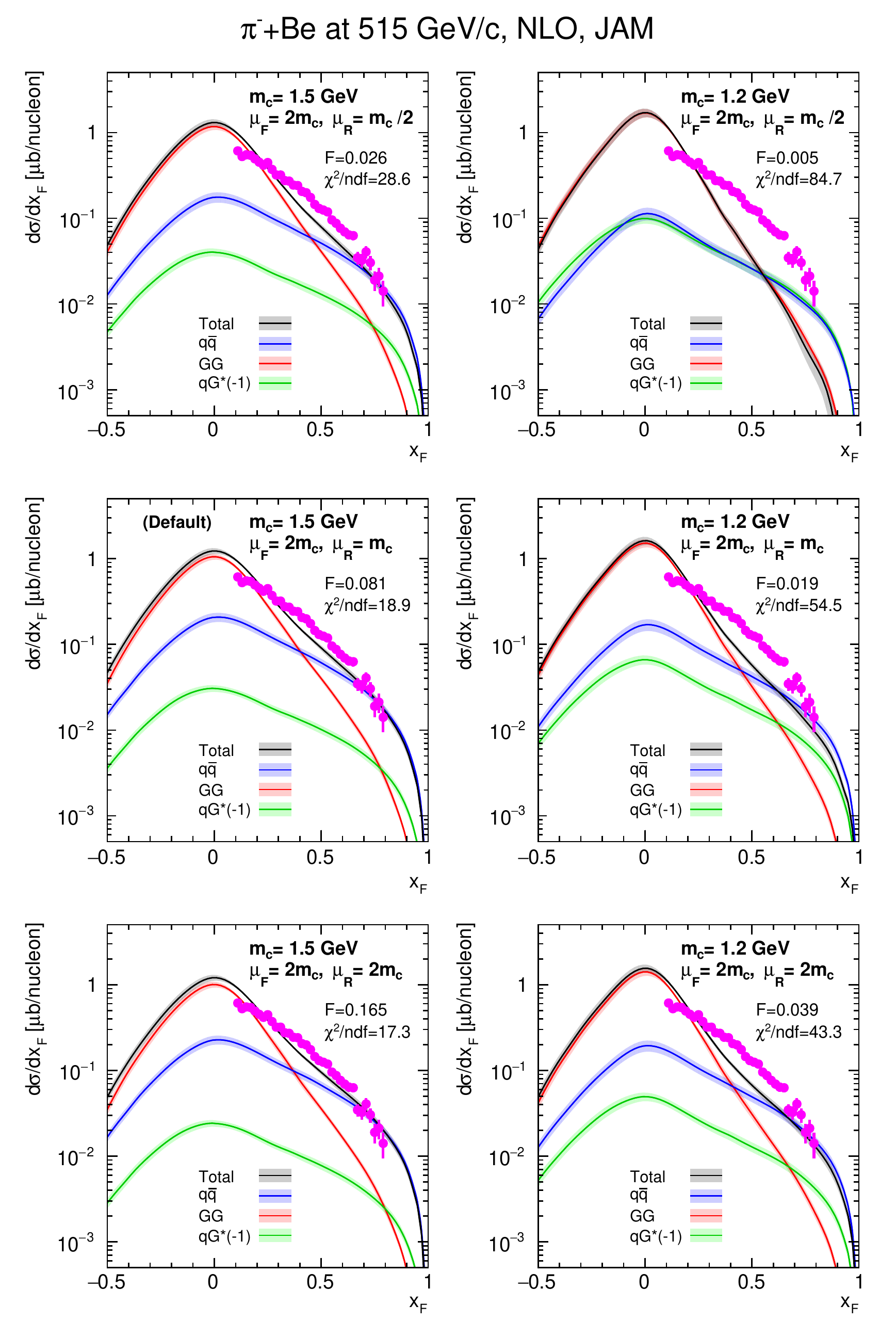}
\caption[\protect{}]{The same as Fig.~\ref{fig3_sys_GRV} but with the
  input of JAM pion PDFs.}
\label{fig3_sys_JAM}
\end{figure}

Figures ~\ref{fig3_sys_GRV} and ~\ref{fig3_sys_JAM} show the
systematic study of comparing the E672/E706 data and CEM NLO
calculation with GRV and JAM pion PDFs with the variation of $m_c$ and
$\mu_R$. In total there are six settings of parameters under
investigation. Overall, the charm quark mass $m_c$ plays a more visible
role than the renormalization scale $\mu_R$ in the systematic
effect. With a smaller charm quark mass $m_c$, the fractions of $q
\bar{q}$ decrease while the fractions of $GG$ increase. The
hadronization factor $F$ drops with the decrease of $m_c$, in
accordance with a large phase space of $c \bar{c}$ production in
Eq.~(\ref{eq:eq1}). The variation of the renormalization scale $\mu_R$
shows a similar but much less significant trend.

For this dataset at the largest beam momentum of 515 GeV/$c$, the $GG$
contribution is dominating in the CEM NLO calculation. A reduction of
$m_c$ from 1.5 to 1.2 GeV/$c^2$ reduces the relative contribution of
$q \bar{q}$ and leads to a deterioration of $\chi^2$/ndf for both GRV
and JAM. Nevertheless, this effect is particularly significant in the
case of JAM. With a reduction of the $q \bar{q}$ contribution, the
large-$x$ gluon density of JAM PDFs is not strong enough to sustain
enough $GG$ contribution in accounting for the cross sections at large
$x_F$. The information of $F$ factor and $\chi^2$/ndf for the
systematic study of this dataset with the SMRS, GRV, xFitter, and JAM
pion PDFs, is shown in Table~\ref{tab:sys_data1}.

\begin{table}[htbp]   
\centering
\begin{tabular}{|cc|c|c|c|c|c|c|c|c|}
\hline
\multicolumn{2}{|c|}{Setting} & \multicolumn{4}{|c|}{$F$}& \multicolumn{4}{|c|}{$\chi^2$/ndf} \\
\cline{3-10}
$m_c$ & $\frac{\mu_R}{m_c}$  & SMRS & GRV & xFitter & JAM & SMRS & GRV & xFitter & JAM \\
\hline
 1.2 & 0.5 & 0.002 & 0.002 & 0.004 & 0.004 &   3.2 &   6.8 &  20.8 &  75.0 \\
 1.2 & 1.0 & 0.010 & 0.009 & 0.014 & 0.017 &   2.1 &   5.0 &  12.7 &  50.6 \\
 1.2 & 2.0 & 0.020 & 0.020 & 0.030 & 0.035 &   1.7 &   4.3 &   9.8 &  40.8 \\
\hline
 1.5 & 0.5 & 0.013 & 0.013 & 0.019 & 0.024 &   1.4 &   2.4 &   7.2 &  28.4 \\
 1.5 & 1.0 & 0.040 & 0.040 & 0.059 & 0.075 &   1.1 &   2.1 &   5.5 &  19.9 \\
 1.5 & 2.0 & 0.084 & 0.084 & 0.121 & 0.153 &   1.0 &   1.9 &   4.1 &  17.6 \\
\hline
\end{tabular}
\caption {Results of $F$ factor and $\chi^2$/ndf value of the best fit
  of the CEM calculations for SMRS, GRV, xFitter and JAM pion PDFs to
  the data of \Jpsi~ production off the beryllium target with a
  515-GeV/$c$ $\pi^-$ beam~\cite{jpsi_data2and3}, with the systematic
  variation of charm quark mass $m_c$ between 1.2 and 1.5 GeV/$c^2$,
  and renormalization scale $\mu_R$ at 0.5, 1.0, and 2.0 $m_c$.}
\label{tab:sys_data1}
\end{table}

From the systematic study of all datasets, the NLO CEM results clearly
favor SMRS and GRV, especially at high energies. The $\chi^2$/ndf,
representing the performance of data description, strongly correlates
with how large the magnitude of gluon density is in the valence
region. As shown in Fig.~\ref{fig_pdf}(c), SMRS and GRV have a
significantly larger gluon density at large $x$ than xFitter and
JAM. Overall, our studies indicate that high-energy \Jpsi~ data have
an increased sensitivity to the pion large-$x$ gluon density in the
NLO calculations, resulting from the enhanced importance of the $GG$
contribution. On the other hand, the relatively small difference in
the valence-quark distributions for various PDFs plays a minor role in
\Jpsi~ production if away from the threshold region, as seen in the
comparison of results of SMRS and GRV.




\section{Discussion}
\label{sec:discussion}

 
From the early CEM LO studies~\cite{Gluck:1977zm, Barger:1980mg}, it
was known that the fixed-target \Jpsi~ production is sensitive to pion
valence-quark distribution at low energies via the $q \bar{q}$
mechanism and to the $GG$ contribution at high
energies~\cite{Gluck:1977zm}. In this study, we confirm the
sensitivity of the fixed-target \Jpsi~ data to the pion's gluon
density in the valence-quark region within the CEM. Moreover, we show
that this sensitivity is further enhanced when the NLO calculations
are performed. The hadronization factors $F$ with the default
parameter setting are found in the range of 0.04--0.08, consistent
with the ones determined from the fixed-target proton-induced \Jpsi~
data~\cite{Gavai:1994in}. This finding supports the usage of the CEM
approach to access the pion PDFs from \Jpsi~ production.

The NLO CEM calculations suggest that the $x_F$ distributions for
fixed-target \Jpsi~ production can serve as a tool for accessing the
pion partonic densities. At low energies the data are predominantly
sensitive to the pion's valence-quark distributions, while at high
energies the data become increasingly sensitive to the gluon
distributions in the pion. Thus, a global fit taking into account the
\Jpsi~ data across a broad energy range is expected to be helpful in
pinning down the large-$x$ gluon density of pions better. At low
energies where the $q \bar{q}$ mechanism dominates, the pion-induced
\Jpsi~ production, having much larger cross sections than the
Drell-Yan process, could be a powerful alternative to the Drell-Yan
process in probing the quark distributions of pions.

We note that the recent effort to include leading neutron DIS data in
the JAM global analysis has provided new constraints on the pion's sea
and gluon distributions at $x
\sim$0.001--0.1~\cite{Barry:2018ort}. Unfortunately, the existing
leading neutron DIS data are not sensitive to the PDF at $x>0.1$. It
is also important to include the direct-photon production as well as
\Jpsi~ production data in the future global fits to place stringent
constraints on the gluon distributions at large $x$. As shown in this
study, the JAM gluon density at large $x$ is too low to reproduce the
\Jpsi~ data. The upcoming tagged DIS experiment at the Jefferson Lab
will be able to extend the sensitive region up to $x=0.2$~\cite{TDIS}.

\section{Summary}
\label{sec:summary}

We have examined the available pion PDFs extracted from the global fit
to Drell-Yan, prompt-photon production or leading neutron DIS
data. These PDFs present pronounced differences, particularly in the
gluon distributions. We have calculated their total and $x_F$
differential cross sections for pion-induced \Jpsi~ production using
the CEM framework at NLO. The calculations are compared to the data
using hydrogen and light nuclear targets.

We observe the importance of the gluon fusion process in \Jpsi~
production, especially at high (fixed-target) energies. We find that
this dominance is even more pronounced in the NLO calculation. Since
the calculated shapes of $x_F$ distributions of $GG$ and $q \bar{q}$
contributions are directly related to the parton $x$ distributions of
corresponding PDFs, a proper description of \Jpsi~ production data,
especially for $x_F>0.5$, imposes a strong constraint on the relevant
pion's parton PDFs. Among the four pion PDFs examined, the CEM NLO
calculations favor SMRS and GRV PDFs whose gluon densities at $x >
0.1$ are higher, compared with xFitter and JAM PDFs. The $GG$
contribution from the latter two pion PDFs drops too fast toward $x_F
= 1$ to describe the data.

Within the CEM, our study clearly indicates that the fixed-target
pion-induced \Jpsi~ data could be useful in constraining the pion
gluon density, particularly at the large-$x$ region. It will be
interesting to perform similar studies using the more sophisticated
NRQCD approach. In the near future, new measurements of Drell-Yan as
well as \Jpsi~ data in $\pi A$ reactions will be available from the
CERN COMPASS experiment. While further theoretical efforts are
required to reduce the model dependence in describing the \Jpsi~
production, we believe that it is important to include the existing
large amount of pion-induced \Jpsi~ data as well as the new ones in
future pion global analysis.

\section*{Acknowledgments}
\label{sec:acknowledgments}

We thank Nobuo Sato and Ivan Novikov for providing us with
LHAPDF6-compatible grid files of JAM and xFitter PDFs. This work was
supported in part by the U.S. National Science Foundation and the
Ministry of Science and Technology of Taiwan.


\end{document}